\pdfoutput=1

\documentclass[prl,superscriptaddress,amsmath,amssymb,twocolumn,nofootinbib]{revtex4-2}

\usepackage{graphicx,bm,amsmath}
\usepackage[usenames,dvipsnames]{xcolor}
\usepackage[colorlinks,bookmarks=false,citecolor=NavyBlue,linkcolor=Red,urlcolor=blue]{hyperref}
\usepackage{amsthm}
\usepackage[bbgreekl]{mathbbol}
\usepackage{dcolumn}
\usepackage[normalem]{ulem}
\usepackage{braket}
\usepackage{comment}

\usepackage{pdfpages} % include pdfs
\usepackage{pgffor} % for loops
\makeatletter
\AtBeginDocument{\let\LS@rot\@undefined}
\makeatother

\def\supplementfilename{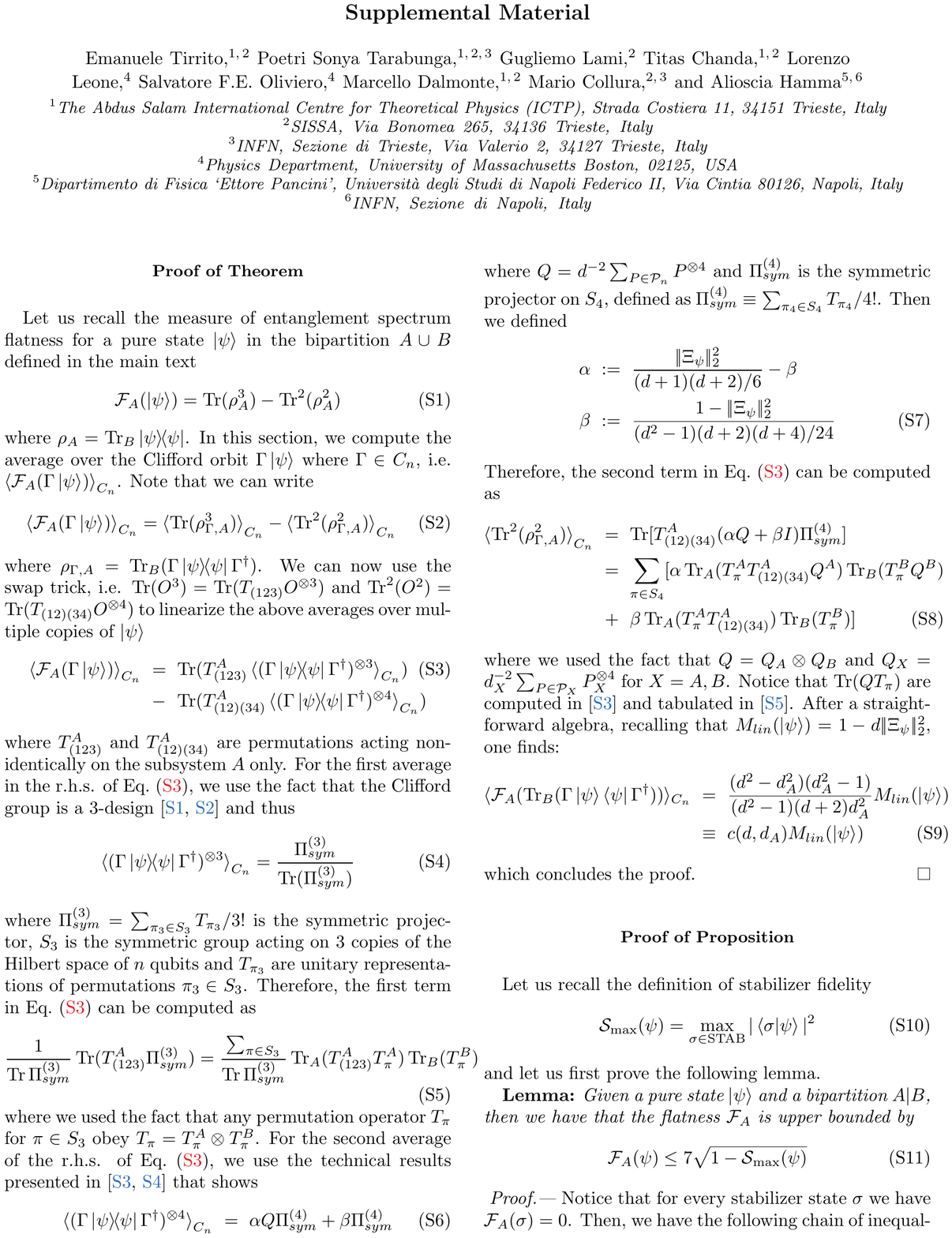}

\pdfximage{\supplementfilename}
\def\numbersupplementpages{\the\pdflastximagepages}

\newcommand{\st}[1]{\ket{#1}\!\!\bra{#1}}
\newcommand{\tr}[0]{\operatorname{Tr}}

\def\doi{http://dx.doi.org/}

\newcommand{\be}{\begin{equation}}
\newcommand{\ee}{\end{equation}}
\newcommand{\bec}{\begin{equation*}}
\newcommand{\eec}{\end{equation*}}
\newcommand{\bea}{\begin{eqnarray}}
\newcommand{\eea}{\end{eqnarray}}

\def\norm#1{\Vert #1\Vert}

\newcommand{\titleinfo}{Quantifying non-stabilizerness through entanglement spectrum flatness}

\newcommand{\Tr}{\text{Tr}}   % Trace 
\begin{document}

%%%%%%%%%%%%%%%%%%%%%%%%%%%%%%%%%%%%%%%%%%%%%%%%%%%%%%%%
\title{\titleinfo}
%%%%%%%%%%%%%%%%%%%%%%%%%%%%%%%%%%%%%%%%%%%%%%%%%%%%%%%%
\author{Emanuele Tirrito}
\affiliation{The Abdus Salam International Centre for Theoretical Physics (ICTP), Strada Costiera 11, 34151 Trieste, Italy}
\affiliation{SISSA, Via Bonomea 265, 34136 Trieste, Italy}
\author{Poetri Sonya Tarabunga}
\affiliation{The Abdus Salam International Centre for Theoretical Physics (ICTP), Strada Costiera 11, 34151 Trieste, Italy}
\affiliation{SISSA, Via Bonomea 265, 34136 Trieste, Italy}
\affiliation{INFN, Sezione di Trieste, Via Valerio 2, 34127 Trieste, Italy}
\author{Gugliemo Lami}
\affiliation{SISSA, Via Bonomea 265, 34136 Trieste, Italy}
\author{Titas Chanda}
\affiliation{The Abdus Salam International Centre for Theoretical Physics (ICTP), Strada Costiera 11, 34151 Trieste, Italy}
\affiliation{Department of Physics, Indian Institute of Technology Indore, Khandwa Road, Simrol, Indore 453552, India}
\author{Lorenzo Leone}
\author{Salvatore F.E. Oliviero}
\affiliation{Physics Department, University of Massachusetts Boston, 02125, USA}
\author{Marcello Dalmonte}
\affiliation{The Abdus Salam International Centre for Theoretical Physics (ICTP), Strada Costiera 11, 34151 Trieste, Italy}
\affiliation{SISSA, Via Bonomea 265, 34136 Trieste, Italy}
\author{Mario Collura}
\affiliation{SISSA, Via Bonomea 265, 34136 Trieste, Italy}
\affiliation{INFN, Sezione di Trieste, Via Valerio 2, 34127 Trieste, Italy}
\author{Alioscia Hamma}
\affiliation{Dipartimento di Fisica ‘Ettore Pancini’, Università degli Studi di Napoli Federico II, Via Cintia 80126, Napoli, Italy}
\affiliation{INFN, Sezione di Napoli, Italy}

\begin{abstract}
Non-stabilizerness - also colloquially referred to as magic - is a resource for advantage in quantum computing and lies in the access to non-Clifford operations. Developing a comprehensive understanding of how non-stabilizerness can be quantified and how it relates to other quantum resources is crucial for studying and characterizing the origin of quantum complexity. 
In this work, we establish a direct connection between non-stabilizerness and entanglement spectrum flatness for a pure quantum state. We show that this connection can be exploited to efficiently probe non-stabilizerness  even in the presence of noise. Our results reveal a direct connection between non-stabilizerness and entanglement response, and define a clear experimental protocol to probe non-stabilizerness in cold atom and solid-state platforms.
\end{abstract}

\maketitle

{\em Introduction.---} 
Simulating quantum states is in general very hard for classical computers. It is expected that the exact classical simulation of arbitrary quantum systems is inefficient, as the resource overhead exponentially grows with the size of the system~\cite{shor1999polynomial}.
For this reason, Feynman put forward the notion of a quantum computer \cite{feynman1986quantum} as only a quantum device would be able to simulate a generic quantum system efficiently, as later proven in~\cite{Lloyd1996simulators}.

Entanglement - one of the  defining characteristics and essential resources for quantum processing and quantum technology - has been thoroughly studied~\cite{amico2008entanglement,eisert2010colloquium,cirac2012goals,bloch2012quantum,aspuru2012photonic,houck2012chip,vandersypen2005nmr,plenio2014introduction,chamon_quantum_2022}. However, probing entanglement is insufficient for quantum advantage, nor it is enough to characterize entanglement by a single number~\cite{vidal2003efficient,van2007classical,van2008completeness,van2013universal,de2009completeness}. For instance, one can obtain highly entangled states by Clifford circuits~\cite{nielsen2002quantum} that can be efficiently simulated classically~\cite{gottesman1997stabilizer,gottesman1998theory,aaronson2004improved,veitch2014resource}; moreover, entanglement complexity is revealed in the finer structure of the entanglement spectrum statistics\cite{chamon2014EmergentIrreversibilityEntanglement,shaffer2014IrreversibilityEntanglementSpectrum,yang2017EntanglementComplexityQuantum}. Transitions between different classes of entanglement complexity are driven by non-Clifford resources~\cite{zhou2020SingleGateClifford,leone2021QuantumChaosQuantum,oliviero2021TransitionsEntanglementComplexity,true2022TransitionsEntanglementComplexity,piemontese2022EntanglementComplexityRokhsarKivelsonsign,leone2020isospectral}, which are indeed also the necessary resources to quantum advantage\cite{gottesman1998heisenberg}.

Any extension of Clifford circuits (that is, circuits only containing  the Hadamard gate, the $\pi/2$-phase gate, and the controlled-Not gate)enables them to perform universal computation by allowing the input states to include the so-called \textit{magic states} \cite{bravyi2005UniversalQuantumComputation,campbell2017roads,bravyi2012magic}.
The canonical magic-state is $|T\rangle=\left( |0\rangle +e^{i\pi/4} |1\rangle \right)/ \sqrt{2}$ which enables the application of a single-qubit unitary $T=\rm{diag}(1,e^{i\pi/4})$. A circuit composed of elements from the Clifford and $T$ gate set acting on the standard computation basis input suffices for universal quantum computation.
Classical simulation of such circuits needs a  run-time scaling exponentially with the number of input magic-state qubits, yet still polynomial in the number of stabilizer-state qubits~\cite{aaronson2004improved}: 
consequently, one can perform an efficient classical simulation for any class of circuits that is nearly stabilizer in the sense that they use only logarithmically many input magic-state qubits \cite{bravyi2016improved,bravyi2016trading,garcia2014geometry}.

\begin{figure*}[t!]
  \centering
    \includegraphics[width=1.\linewidth]{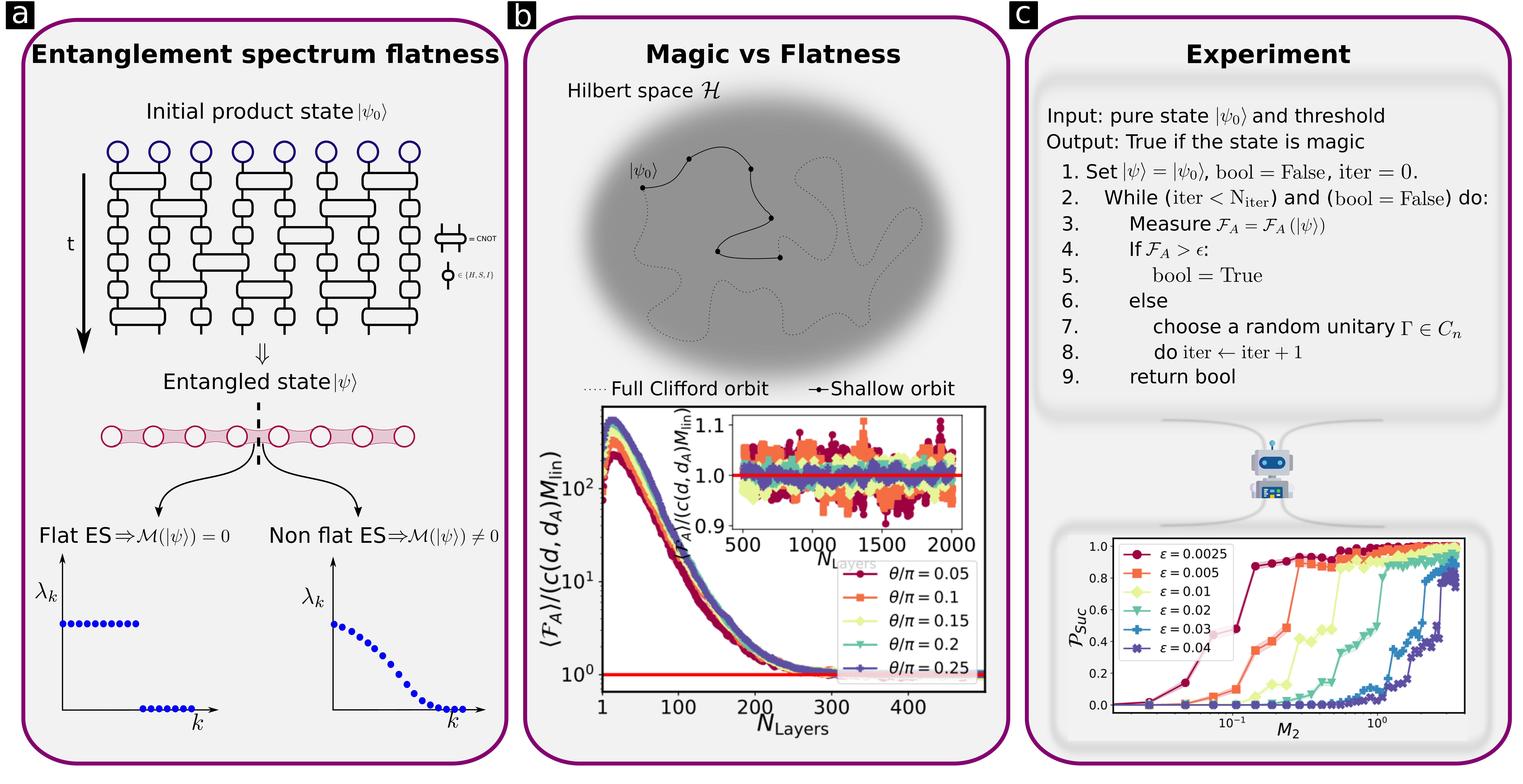}
\caption{\label{fig:fig_1} \textbf{Summary of the results:} (a) A schematic of the method to quantify the non-stabilizerness of a pure state. We start from a product state and then we apply random Clifford gates (both single and two-qubit gates, see text). After $N_{\rm{Layers}}$, the state is fully entangled. Checking the 
entanglement spectrum in any bipartition, we can distinguish whether the initial state possesses non-stabilizerness. 
(b) In the upper panel, a sketch of the Clifford orbit of a pure state is shown. 
In the lower panel, we show the relation between anti-flatness $\mathcal{F}_A$ and non-stabilizerness, quantified by $c(d,d_A) M_{\rm{lin}} $. We initialize the system as a product state $|\psi(0) \rangle=\otimes^{n}_{i=1} |\psi_i \rangle$, where $|\psi_i\rangle=\frac{1}{\sqrt{2}} \left( |0\rangle +e^{i\theta}|1\rangle \right)$ and $n=14$, for different values of $\theta$. Then we apply several random Clifford layers $N_{\rm{Layers}}$. In the limit of a very deep circuit, the ratio approaches $1$ as predicted by the theorem as shown in the inset.
(c) Algorithm to determine if a state is a stabilizer state or possesses some non-stabilizerness. We show the pseudocode in the upper part of the panel. In the lower part, we show the probability of catching a non-stabilizerness.  We generate product states $|\psi(0) \rangle=\otimes^{n}_{i=1} |\psi_i \rangle$,  where $|\psi_i\rangle=\frac{1}{\sqrt{2}} \left(|0\rangle +e^{i\theta}|1\rangle \right)$ for $n=12$ qubits, and, we fix the number of Clifford layers $N_{\rm{Layers}}=100$.  After performing $N_R=1000$ realizations, we compute the probability of success $\mathcal{P}_{\rm{suc}}$ for different values of threshold, as a function of the initial value of non-stabilizerness in the initial state calculated using the second SRE. 
}
\end{figure*}

As explained above, non-Clifford resources are necessary for any quantum advantage so they are a precious resource in quantum information processing, often dubbed non-stabilizerness, or, more colloquially, magic~\cite{bravyi2012magic}. The resource theory for non-stabilizerness has been developed in the last few years. Proposed measures of magic often incorporate the concept of quasi-probability, with many of these measures relying on the discrete Wigner formalism as a foundation\cite{wigner1997quantum,gross2007non,veitch2012negative,wootters1987wigner,hudson1974wigner,buvzek1992superpositions}.
Prominent examples include the relative entropy of magic and the mana(for qudit) \cite{veitch2014resource}, the stabilizer rank \cite{bravyi2016trading,bravyi2016improved,bravyi2019simulation}, the robustness of magic \cite{howard2017robustness,heinrich2019robustness}, the thauma(for qudit) \cite{wang2020efficiently}, 
and the stabilizer extent \cite{heimendahl2021stabilizer}. Most of these quantifiers of non-stabilizerness are difficult to compute even numerically. One computable and remarkable measure of non-stabilizerness has been introduced recently: the Stabilizer R\'enyi Entropy (SRE) \cite{leone2022stabilizer}, which can be computed efficiently for matrix product states \cite{oliviero2022MagicstateResourceTheory,haug2023quantifying,haug2023stabilizer,tarabunga2023manybody}, is amenable to experimental measurement \cite{haug2022scalable,oliviero2022measuring} and has tight connections with quantum verification and benchmarking~\cite{Leone2023nonstabilizernesshardness}. 

In this letter, we establish a deep connection between the SRE and the flatness of the entanglement spectrum associated to a subsystem density operator. At the conceptual level, this connection shows clearly how this resource is associated with entanglement structure: non-stabilizerness is directly tied to entanglement response, a quantum analogue of 'heat capacity' for thermodynamic systems. At the very same time, it opens the door for important practical applications. We present a simple practical protocol for experimentally probing this quantity efficiently using randomized measurement techniques. 

Our main findings are summarized in Fig.~\ref{fig:fig_1}. In Fig. \ref{fig:fig_1} (a), we show the set-up that we use to make the connection between nonstabilizerness and entanglement response concrete: we prepare initial states as a product states and evolve them using random Clifford gates, followed by the measurement of the entanglement spectrum flatness. We find that a state possesses non-stabilizerness if and only if its entanglement spectrum is not flat. In the second panel (Fig. \ref{fig:fig_1} (b)), we illustrate the Clifford orbit of a pure state: its non-stabilizerness is proportional to its average flatness over the orbit. Finally, in the third panel (Fig. \ref{fig:fig_1} (c)), we present an algorithm for detecting non-stabilizerness and show the probability of success as a function of their degree of non-stabilizerness.

{\em Stabilizer R\'enyi entropy and the flatness of entanglement spectrum.---} 
In this section, we define the SRE and its connection with the flatness of the entanglement spectrum. In particular, we will show that we can quantify the non-stabilizerness of an arbitrary pure state by taking the average of the flatness along its Clifford orbit.

Consider the $d=2^n-$dimensional Hilbert space of $n$ qubits $\mathcal H\simeq \mathbb C^{\otimes 2n}$. A subset $\Lambda$ of the $n$ qubits with $|
\Lambda|=n_\Lambda$  defines a subsystem that is obviously represented by the $d_\Lambda=2^{n_\Lambda}-$dimensional Hilbert space 
$\mathcal H_\Lambda\simeq \mathbb C^{\otimes 2n_\Lambda}$.
Let $P_i \in\{ I,X,Y,Z\}$ 
be the Pauli operators on the $i-$th single qubit space $\mathbb C^2$. 
Pauli operators on the full $\mathcal H$ have the form $P=\otimes_i^n P_i$ and local Pauli operators can be written also as $P_X = \otimes_{i\in \Lambda} P_i$. Let us call
 $\mathcal{P}_n$ the group of all $n$-qubit Pauli operators with phase $1$, and define $\Xi_{\psi} (P)=d^{-1} \tr(P\psi)$ as the squared (normalized) expectation value of $P$ in the pure state $\psi\equiv\st{\psi}$ with $d=2^n$ the dimension of the Hilbert space of $n$ qubits. Moreover, $\Xi_{\psi}$ is the probability of finding $P$ in the representation of the state $\psi$.
Now we can define the SREs as: 
\be \label{eq:SRE_def}
M_{\alpha} \left( \ket{\psi} \right)=E_{\alpha}(\Xi_{\psi})-\log d
\ee
where $E_{\alpha}(\Xi_{\psi})$ is the $\alpha$-R\'enyi entropy of the probability distribution $\Xi_{\psi}$.
The SRE is a good measure from the point of view of resource theory: it tells how many magic states can be distilled, and it is, as such, an important resource for quantum information algorithms\cite{leone2022stabilizer}. The SRE has the following properties: (i) faithfulness $M_{\alpha}\left(|\psi\rangle \right)=0$ iff $|\psi \rangle \in \rm{STAB}$, otherwise $M_{\alpha}(|\psi\rangle) >0$, (ii) stability under Clifford operations: $\forall \Gamma  \in C_n$ we have that $M_{\alpha} \left( \Gamma |\psi \rangle \right) = M_{\alpha} \left(|\psi\rangle \right)$ and (iii) additivity  $M_{\alpha} \left( |\psi \rangle \otimes |\phi \rangle \right) = M_{\alpha} \left( |\psi\rangle \right) + M_{\alpha} \left( |\phi\rangle \right) $ (the proof can be found in \cite{leone2022stabilizer}). 

Another useful measure of non-stabilizerness is given by the stabilizer linear entropy, defined as
\be 
M_{\rm{lin}} (|\psi \rangle)=1-d \norm{ \Xi_{\psi} }^2_2 ,
\ee
which obeys the following properties: (i) faithfulness $M_{\rm{lin}}\left(|\psi\rangle \right)=0$ iff $|\psi \rangle \in \rm{STAB}$, otherwise $M_{\rm{lin}}(|\psi\rangle) >0$, (ii) stability under Clifford operations: $\forall  \Gamma  \in C_n$ we have that $M_{\rm{lin}} \left( \Gamma |\psi \rangle \right) = M_{\rm{lin}} \left(|\psi\rangle \right)$ and (iii) upper bound
$M_{\rm{lin}}<1-2\left(d+1\right)^{-1}$.
The relationship between the second SRE $M_2$ and the linear non-stabilizing entropy follows easily from the equation below
\be 
M_2(\ket{\psi}) = -\log \left[ 1-M_{\rm{lin}}(\ket{\psi}) \right] .
\ee

\begin{figure}[t!]
  \centering
    \includegraphics[width=0.93\linewidth]{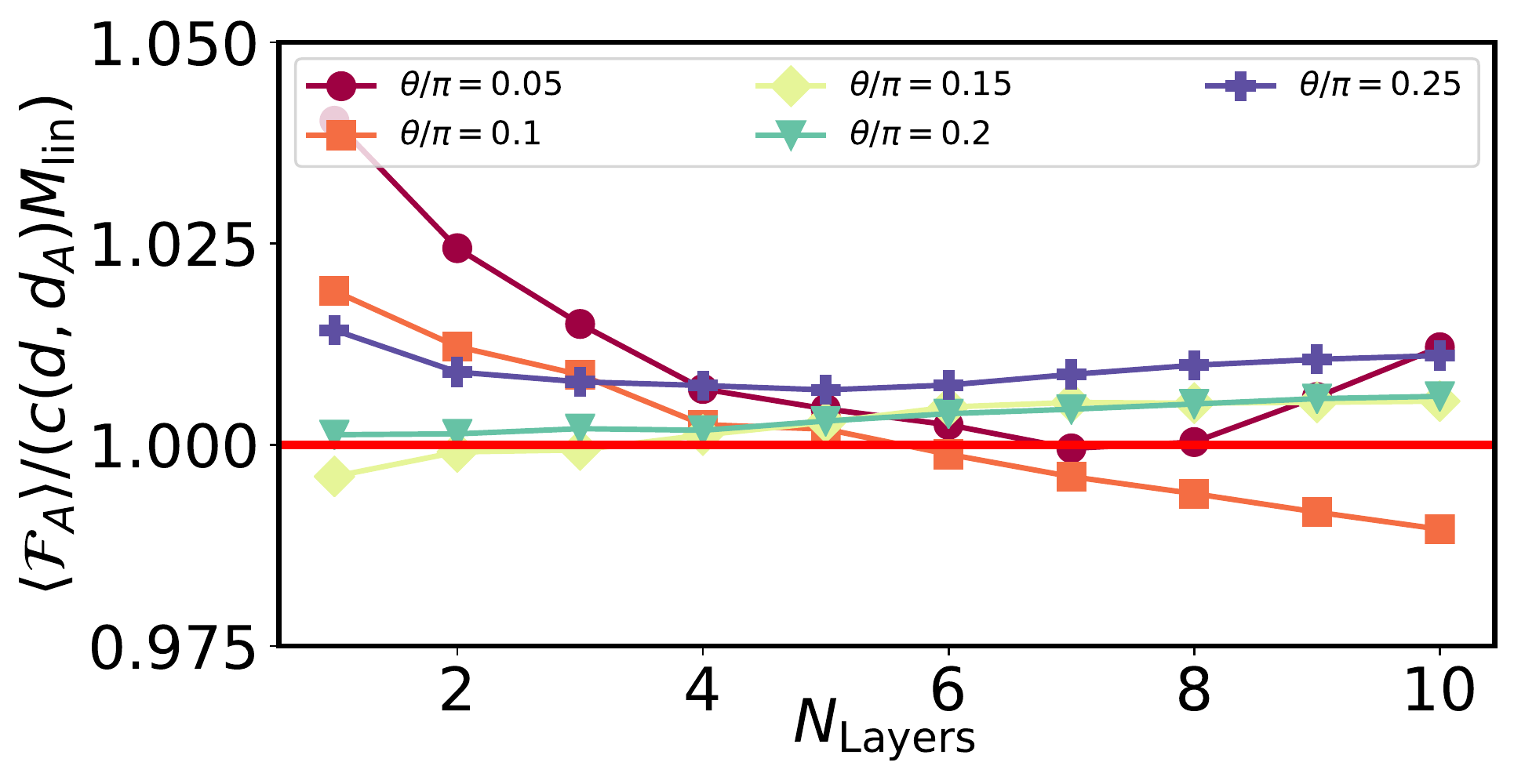}
\caption{\label{fig:fig_2} \textbf{Numerical simulations of  shallow circuits:}  We prepared the initial state in the volume law phase and we plot the ratio $\mathcal{F}_A/c(d,d_A) M_{\rm{lin}}$ as a function of number of Clifford layers $N_{\rm{Layers}}$ of shallow circuit. As shown in the plot, the ratio approaches $1$ very fast verifying Eq.~\eqref{eq:f_vs_min}. }
\end{figure}

Let us now discuss the relationship between the SRE and the flatness of the entanglement spectrum. Consider a pure state $\psi$ in a bipartite system $\mathcal H =\mathcal H_A\otimes \mathcal H_B$ and its reduced density operator  $\psi_A= \Tr_B \psi$. 
The (anti-)flatness of its entanglement spectrum is defined as
\be 
\mathcal{F}_A (\psi) := \Tr \left( \psi^3_A \right) -\Tr^2 \left( \psi^2_A \right)
\ee
One can easily check that $\mathcal{F}_A (\psi)=0$ if the entanglement spectrum is flat, i.e. if the spectrum $\lambda_{\alpha}=1 / \chi$ for some integer $1 \leq \chi \leq \rm{min} \left( d_A,d_B \right) $, whereas $\mathcal{F}_A \left( |\psi \rangle \right) > 0$ in other cases. Notice that in order for $\mathcal{F}_A (\psi)\ne0$
the state must be either not entangled or without any magic. While every linear combination of different moments would be a sensible measure of (anti-)flatness, the one proposed above is the most natural one as it is the variance of the corresponding probability distribution according to itself: if a state $\sigma$ is given in its spectral resolution $\sigma=\sum_i p_i \sigma_i$, then $\mathcal F(\sigma) = \mbox{Var} (\{p_i\}):= \langle (p_i-\langle p\rangle_p)^2\rangle_p$.

In this paper, we use the flatness of the entanglement spectrum to quantify or witness non-stabilizerness of a pure state. 

\textbf{Theorem:} {\em The Stabilizer Linear Entropy $M_{\rm{lin}}$ of a pure state $|\psi\rangle)$ is proportional to the anti-flatness of the entanglement spectrum averaged over the Clifford orbit:
\be \label{eq:f_vs_min} 
\langle \mathcal{F}_A(\Gamma\ket{\psi})  \rangle_{C_n} =c(d,d_A) M_{\rm{lin}} (|\psi\rangle) \, ,
\ee
where $\langle \cdot   \rangle_{C_n}$ denotes the average over the Clifford orbit $\Gamma\ket{\psi}$ and the proportionality constant $c(d,d_A)\sim (d^2-d^2_A)d^{-3}$ for large $d$, see \cite{SeeSupplementalMaterial} for the proof.}

Notice that the above result holds true for any bipartition of the system, which is reflected in the constant $c(d,d_A)$. We see that a pure stabilizer state possesses a flat entanglement spectrum over all its Clifford orbit and anti-flatness is stable under Clifford operations. Moreover, one can utilize a measurement of anti-flatness to measure  $M_{\rm{lin}}$. The theorem above poses also a relationship between entanglement and magic. Indeed, without entanglement, there is no anti-flatness in the reduced density operator. Along the Clifford orbit, entanglement is near-maximal and this is reflected in $c(d,d_A)$. It would be interesting to see whether (for an equal bipartition) anti-flatness assumes the form $ \mathcal{F}_A(\ket{\psi})   \sim g(d_A)\mbox{Pur}^{\beta}(\psi_A) M_{\rm{lin}} (|\psi\rangle)$ or if this relation only holds for highly entangled states.

\begin{figure}[t!]
  \centering
    \includegraphics[width=0.9\linewidth]{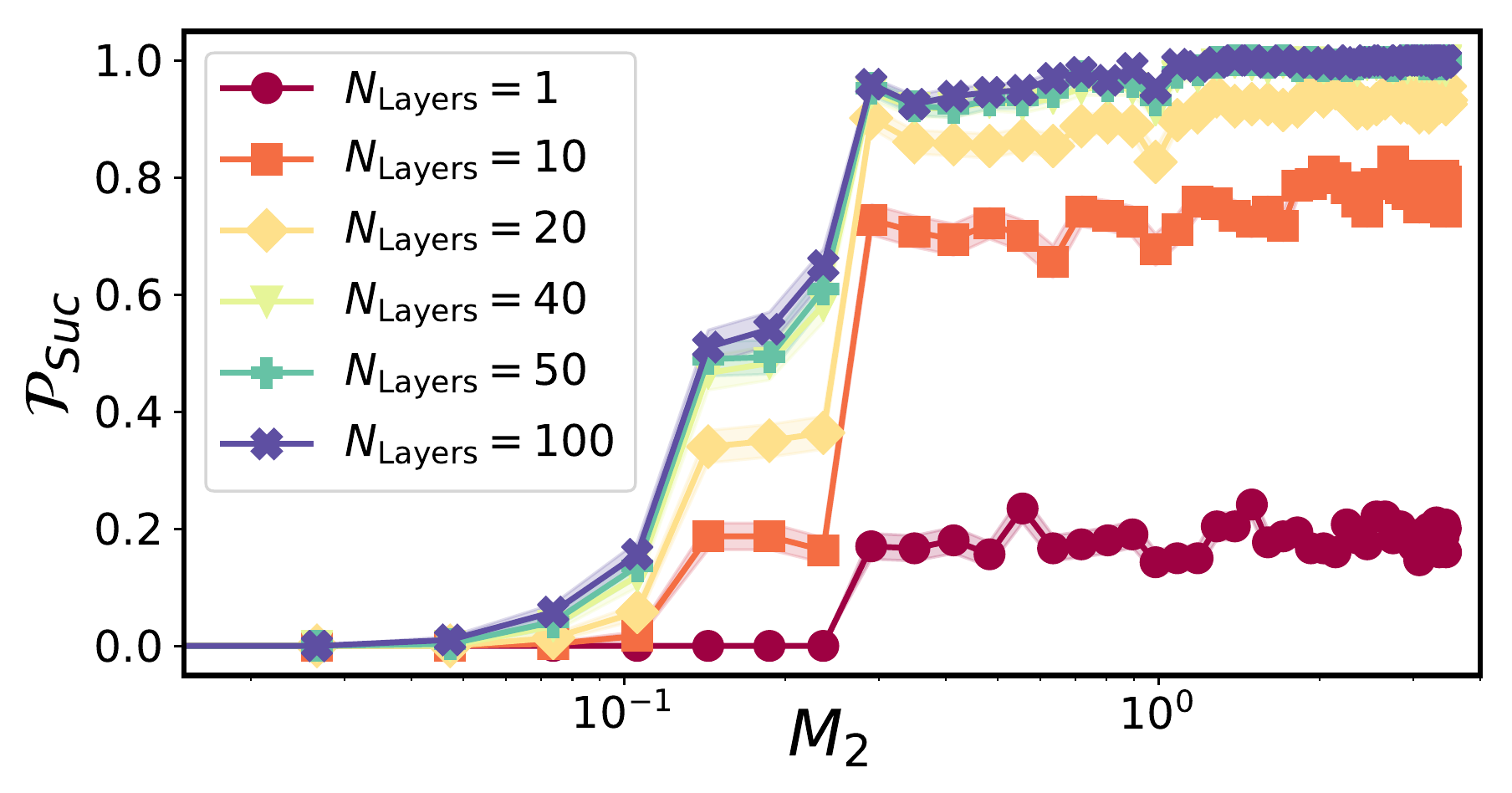}
\caption{\label{fig:fig_3} \textbf{Algorithm sensitivity:} Probability of success $\mathcal{P}_{\rm{Suc}}$, for a fixed threshold $\epsilon=0.005$ and for different $N_{\rm{Layers}}$. After collecting $N_{\text{R}}= 1000$ realizations, we compute the probability of success $\mathcal{P}_{\rm{Suc}}$, as a function of the initial value of non-stabilizerness calculated using the second SRE $M_{2} \left( |\psi_0 \rangle \right)$. }
\end{figure}

\paragraph{Numerical experiments.---} As it was shown in \cite{oliviero2022measuring}, SRE can be experimentally measured via randomized unitaries~\cite{elben2023randomized}, providing an important handle on the quality of a quantum circuit. However, SRE is a very expensive quantity to measure, requiring in general exponential resources (though better than state tomography). The result of the theorem opens the way to a very efficient way to measure SRE. However, things are not so simple. In the best case scenario, $c(d,d_A) = O(d^{-1})$, which means that one needs to resolve an exponentially small quantity, thereby requiring again exponential resources - even if with the considerable advantage that operations on a small subset are needed, thus relaxing one of the most challenging requirements of previous methods. This is because $\psi_A$ is typically very entangled over $C_n$ and therefore $\mathcal F_A$ is very close to be flat. Another issue is that, for weakly entangled states, direct exploitation of the theorem is extremely challenging in practice, as we shall demonstrate numerically in the following. One can intuitively understand that as for very weakly entangled states there are very few eigenvalues at all in the entanglement spectrum. As an extreme example, the entanglement spectrum of a product state is absolutely flat, regardless whether the state possesses any degree of non-stabilizerness. A very long circuit (inevitably, very sensitive to noise) will thus be required in those cases. 

The key insight is that we can get around the requirement of a full Clifford orbit by (numerically) analyzing the intermediate regime. Approaching volume law one might be able to see a deviation from a flat spectrum without having to resolve an exponentially small quantity. If this is true, one would have found a witness for non-stabilizerness that is efficiently computable and measurable. Moreover, as one gets into the volume law for the entanglement phase, one should be able to evaluate accurately the actual value of $M_{lin}$, even without averaging over all the Clifford orbit. Of course, in this case, one still needs to resolve a very small quantity.

We consider an initial state that is a product state of $n$ qubits with linear topology $|\psi_0 \rangle=\otimes^{n}_{i=1} |\psi_i \rangle$, where $|\psi_i\rangle=\frac{1}{\sqrt{2}} \left( |0\rangle +e^{i\theta}|1\rangle \right)$. This state has initially computable non-stabilizerness (vanishing for $\theta=0,\pi/2$). Note that $\theta=\pi/4$ corresponds to the canonical T-state. The state $|\psi_0 \rangle$ is then evolved under a random Clifford circuit of depth $N_{Layers}$ denoted by $U_{\rm{Cl}}=\prod_k^{N_{Layers}} U_k$, where $U_k$ contains $n-1$ Clifford gates (Hadamard, phase $e^{i\pi/2}$ gate and CNOT)\cite{nielsen2002quantum} between nearest neighbors.

We are interested in how the entanglement spectrum, that is,  the eigenvalues of the reduced density matrix $\rho_A=\Tr_B |\psi\rangle \langle \psi |$ (for $d_A=d_B=2^{n/2}$ and $n=14$ qubits) evolves under random Clifford circuit evolution. %{\bf AH uniform $n,N$ notation}
In Fig.~\ref{fig:fig_1} (b), we present the average of the anti-flatness $\langle \mathcal{F}_A \rangle$ as a function of the circuit depth $N_{\rm{Layers}}$.
The average is obtained from $N_R=1000$ different realizations and it is calculated for various values of $\theta$. For a small number of Clifford layers, the anti-flatness increases and exhibits a sharp dependence on $\theta$. When the circuit is very deep, the system explores a very large portion of its Clifford orbit, and the ratio between average anti-flatness $\langle \mathcal{F}_A \rangle$ and $c(2^n,2^{n/2}) M_{\rm{lin}}$ approaches 1 (the solid red line in the inset of Figure \ref{fig:fig_1}), as predicted by the Theorem.

In Fig.~\ref{fig:fig_2}, we show that one can accurately estimate $M_{lin}$ even by shallow Clifford circuits provided one starts with volume law entanglement.  We again consider a $n=14$ qubit system in a volume law phase by subjecting the initial state $| \psi_0 \rangle$ to $N_{\rm{Layers}}=1500$ Clifford layers, for various values of $\theta$. We then plot the ratio $\langle \mathcal{F}_A \rangle / c(2^{n},2^{n/2}) M_{\rm{lin}}$ as a function of the number of Clifford $N_{\rm{Layers}}$. The theoretical line predicted by the theorem is shown as a solid red line. Notably, we observe that even for circuits as short as $N_{\rm{Layers}}=7$ Clifford layers, the average anti-flatness reaches the value predicted by the theorem~\cite{inprep}.

\paragraph{Probing non-stabilizerness through flatness. ---} As we discussed above, one could probe non-stabilizerness by probing flatness, which is amenable to be measured in experiments \cite{pichler2016measurement,johri2017entanglement,choo2018measurement}. However, a na\"ive application of the theorem would result in a very costly procedure. We present an algorithm  
that can efficiently witness magic by exploring the Clifford orbit in the intermediate region between weak and volume-law entanglement~\cite{inprep}. Since measuring non-stabilizerness can be resource-intensive, the concept of witness provides a scalable approach to assess the accurate implementation of stabilizer operations or evaluate the fit of quantum hardware for preparing magic states.

The procedure works as follows: (1) Start with $|\psi_0 \rangle$, a pure state. (2) Draw a random  Clifford gate $\Gamma$ and apply it to   the initial state:  $|\psi_\Gamma\rangle\equiv \Gamma |\psi_0 \rangle$. (3) Measure the entanglement spectrum anti-flatness $\mathcal F_A (\psi_\Gamma)$~\footnote{For small partitions, this can be done either via state tomography, or utilizing the random unitary toolbox~\cite{elben2023randomized}}. If the original state $|\psi_0 \rangle$ is a stabilizer state, the output of the circuit is still a stabilizer state with zero anti-flatness.  On the contrary,  if $|\psi_0 \rangle$ has a non-vanishing amount of non-stabilizerness, we expect that even a modest exploration of the Clifford orbit will result into a non-flat entanglement spectrum. Therefore, if after a number of Clifford unitaries 
we measure $\mathcal{F}_A> 0$ we can establish that the initial state possesses non-stabilizerness. The resulting algorithm is summarized in Fig. \ref{fig:fig_1} (c). In this algorithm, we set both the number of iterations (which determines the number of Clifford layers) and the threshold for measuring flatness. 

Notably, our proposed protocol does not demand an exhaustive exploration of the Clifford group, which is exponentially large. Instead, our findings in the previous section demonstrate that a shallow quantum circuit generated by fixing the number of Clifford layers to a reasonably small value is sufficient for detecting non-stabilizerness with a high probability.
This is illustrated in Fig.~\ref{fig:fig_1} (c): we show the probability of success $\mathcal{P}_{\rm{Suc}}$ (for $n=12$ qubits) as a function of the initial value of non-stabilizerness calculated using the second SRE defined in Eq.(\ref{eq:SRE_def}). In order to address the role of errors in the measurement of $\mathcal{F}_A$, we introduce a threshold value $\epsilon$ for our test. The success probability is defined as the number of times in which the algorithm gives True as output, thus detecting the non-stabilizerness of the initial state normalized to the total number of iterations. 

Fig.~\ref{fig:fig_1} panel $(c)$ displays a knee point effect of the probability of success $\mathcal{P}_{\rm{Suc}}$  as a function of the non-stabilizerness calculated using the second SRE $M_2$, depending on the threshold  value $\epsilon$. While, as argued earlier, away from volume law, the general behavior of this algorithm requires a numerical analysis, the knee-point can be explained analytically in a rigorous way:

{\bf Proposition:} {\em Define $\mathcal{S}_{\max}(\psi):=\max_{\sigma}|\braket{\sigma|\psi}|^2$  the stabilizer fidelity with $\sigma$ a stabilizer state. Then, if $\mathcal{S}_{\max}(\psi)>1-\epsilon^2/7$, i.e. if the state is too close to a stabilizer state, the success probability is zero, that is,  $\mathcal{P}_{\rm{Suc}}(\epsilon)=0$. }

The above proposition provides insight into the sensitivity of the algorithm shown in Fig.~\ref{fig:fig_1} with respect to the stabilizer fidelity $\mathcal{S}_{\max}$, which is closely linked to the stabilizer entropy $M_{\rm{lin}}$. See \cite{SeeSupplementalMaterial} for the proof.

In Fig. \ref{fig:fig_3}, we present the probability of success $\mathcal{P}_{\rm{Suc}}$ (for $n=12$ qubits) for a different maximum number of Clifford layers $N_{\text{Layers}}$.
We fix the threshold $\epsilon = 0.005$ and we compute the probability as a function of non-stabilizerness calculated by $M_2 \left( |\psi_0 \rangle \right)$ of the initial state. The plot shows that increasing the number of algorithmic iterations $N_{\text{Layers}}$ push the probability of success to $1$ for any fixed values of non-stabilizerness.

\begin{figure}[t!]
  \centering
    \includegraphics[width=0.5\linewidth]{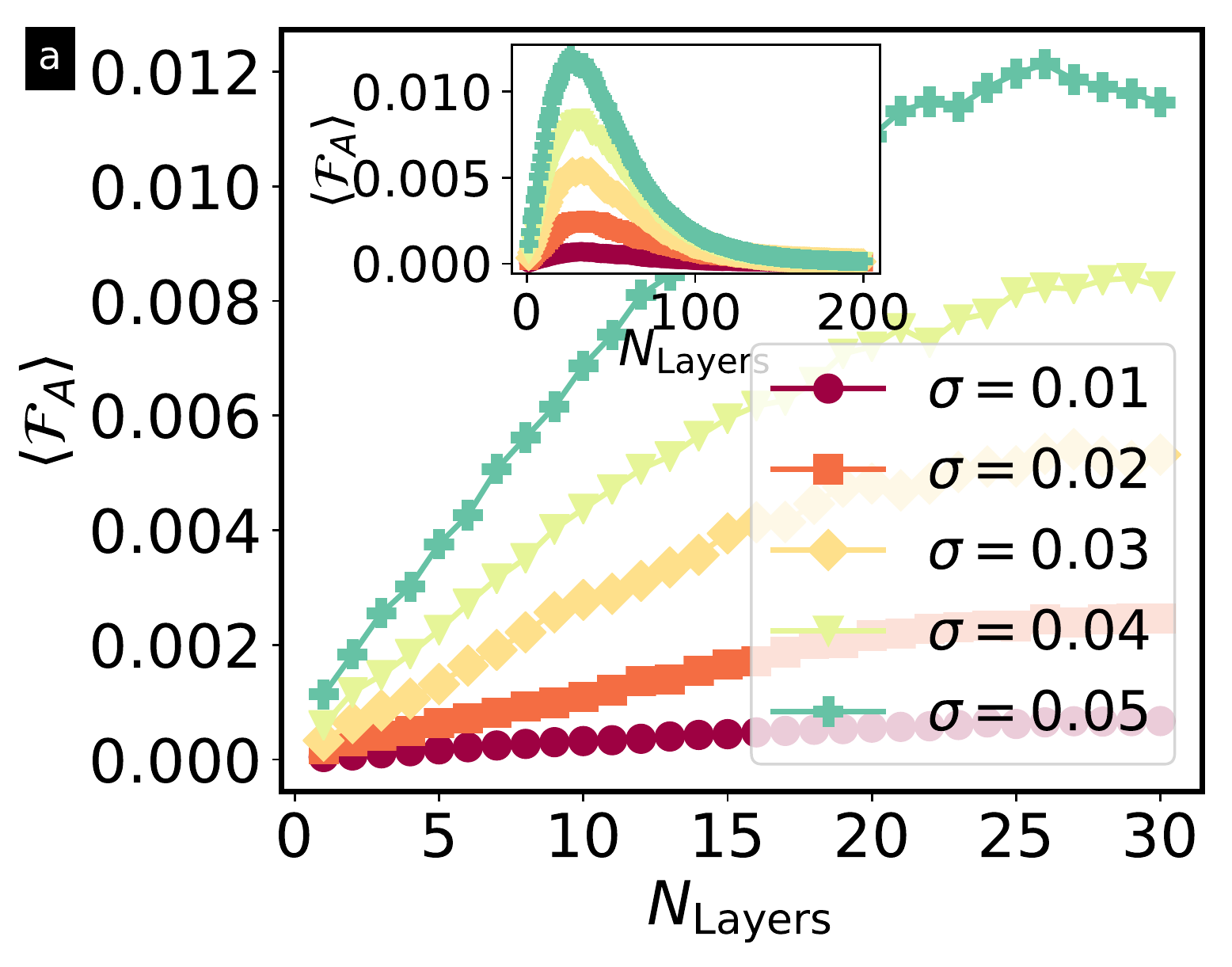}\includegraphics[width=0.52\linewidth]{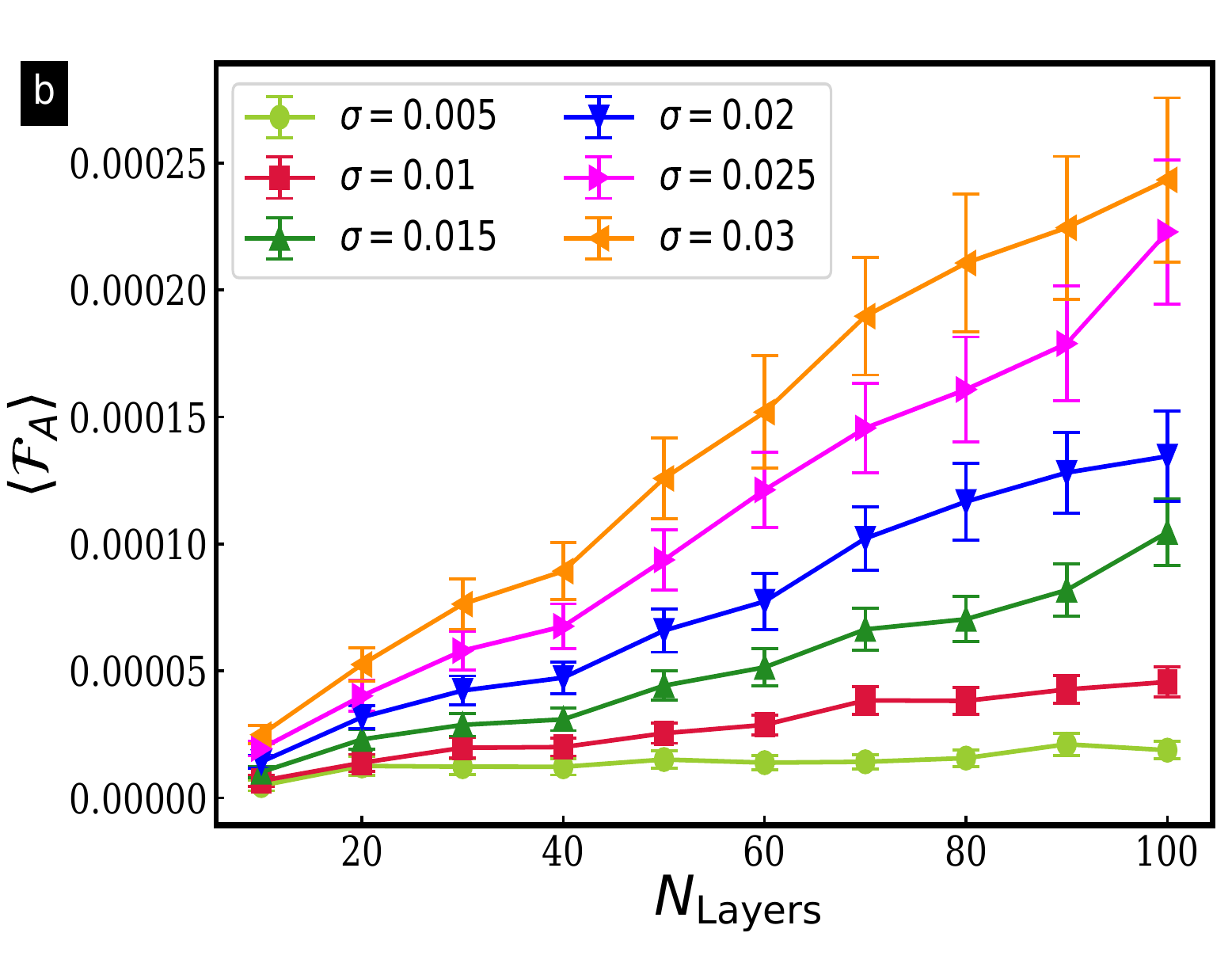}
\caption{\label{fig:fig_4} \textbf{Flatness in noisy circuit:} (a) We show the average of the anti-flatness, over $N_{\rm{R}}=1000$ realizations, as a function of $N_{\rm{Layers}}$. We start from  an initial STAB state $|\psi_0 \rangle=\frac{1}{\sqrt{2}} \left( |0\rangle + |1\rangle \right)$. We inject magic using a modified CNOT gate and the average of the anti-flatness $\mathcal{F}_A$ increases after few layers of our circuit.  (b) We show the average of flatness, over $N_{\rm{R}}=1000$ realizations, as a function $N_{\rm{Layers}}$. We initialize the system in the ground state of Toric code, that is a STAB state, on a $4 \times 2$ unit cell (16 spins). We show that the anti-flatness increases almost linearly with the number of Clifford layers $N_{\rm{Layers}}$. Error bars correspond to a $95\%$ confidence interval.}
\end{figure}

\paragraph{Noisy Clifford circuit.---} So far we assumed that Clifford unitaries are ideal. In reality, they have a residual noise due to the fact that Clifford circuits are fine-tuned. In this situation, it is more natural to perform error mitigation at the level of channels rather than states. Consider a simple error model where each two-qubit Clifford $U (k)$ is independently affected by unitary noise. In particular, every two-qubit gate is transformed as follows:
\be \label{eq:noisy_gate}
\tilde{U}(k)=e^{-i \sum_{\alpha} \epsilon_{\alpha} P^{\alpha}} U(k) e^{i \sum_{\alpha} \epsilon_{\alpha} P^{\alpha}}
\ee
where $\epsilon_{\alpha}$ is a random number chosen from a Gaussian distribution with average zero and standard deviation $\sigma$ that represents here the strength of the noise (see \cite{SeeSupplementalMaterial} for more details). The choice of coherent noise is due to the fact that SRE is a proper measure of distillable magic only for pure states. For mixed states, it still has an important operational meaning in quantifying resources beyond Clifford~\cite{leone2022stabilizer}: for example, it is the key resource for establishing the cost of direct fidelity estimation~\cite{Leone2023nonstabilizernesshardness}, cleansing algorithms and efficient purity estimation~\cite{leone2023phase}. The study of the effect of incoherent noise is to be carried out in future work. 

Introducing noise to Clifford gates represents a magic-state injection that
can be accurately captured by measuring the anti-flatness $\mathcal{F}_A$. 
In Fig. \ref{fig:fig_4} (a) we present the evolution of the average of the anti-flatness $\mathcal{F}_A$ for a noisy Clifford circuit with $n = 14$ qubits. We initialize our system in a stabilizer state 
$|\psi\rangle=\frac{1}{\sqrt{2}} \left( |0\rangle +|1\rangle \right)$ and then we measure the anti-flatness after every Clifford layer.   
Moreover, we also investigate the effect of noise starting from the ground state of the toric code - a stabilizer code formulated on a square lattice \cite{kitaev2003fault,dennis2002topological,raussendorf2007topological,fowler2012surface}. The basic construction of the toric code is a square lattice with a spin-$1/2$ degree of freedom on every bond, the physical qubits.
The model is given in terms of a Hamiltonian $\hat{H}=-\sum_{\nu} A_{\nu}-\sum_{p} B_{p}$, where $p$ runs over all plaquettes and $\nu$ over all vertices (sites). The ground state of the toric code is a stabilizer state of the sets $\left \lbrace A_{\nu} \right \rbrace$ and $\left \lbrace B_{p} \right \rbrace$. After applying a Clifford circuit with a transformed CNOT gate, we measure the anti-flatness $\mathcal{F}_A$ after every layer. 
In Fig. \ref{fig:fig_4} (b) we show the evolution of $\mathcal{F}_A$ for different strengths of noisy $\sigma$. It increases almost linearly with the number of Clifford layers. These results quantify how, upon close inspection of the microscopic imperfections, it is possible to define an error threshold that is able to discriminate between magic injected by errors along the Clifford orbit, and intrinsic magic of the original state. 

\paragraph{Conclusions.---} We have demonstrated how non-stabilizerness of quantum states, while completely unrelated to entanglement {\it per se}, is deeply and exactly related to entanglement response, via the entanglement spectrum flatness of arbitrary partitions. Leveraging on this connection, we have formulated a simple protocol to efficiently witness and quantify non-stabilizerness in quantum systems, that is applicable to both atom and solid state settings where local operations and probing are available. The protocol is particularly efficient for states with volume law entanglement, and can cope with the inevitable presence of noise, as we demonstrate utilizing both random states and toric code dynamics. Our results pave the way for witnessing non-stabilizerness in large scale experiments - a pivotal step to demonstrate computational advantage -, and motivate further study of non-stabilizerness in quantum many-body systems, in particular, in connection to critical behavior, where entanglement response is expected to be particularly relevant.

\begin{acknowledgments}
\textit{Acknowledgments.---} M.D. thanks V. Savona for insightful discussions. The work of M.D. and P.S.T. was partly supported by the ERC under grant number 758329 (AGEnTh), by the MIUR Programme FARE (MEPH), and by the European Union's Horizon 2020 research and innovation programme under grant agreement No 817482 (Pasquans). P.S.T. acknowledges support from the Simons Foundation through Award 284558FY19 to the ICTP. This work was also supported by the PNRR MUR project PE0000023-NQSTI (M.C., M. D., and A.H). 
A.H., L.L. S.O. acknowledge support from NSF award number 2014000. A.H. acknowledges financial support %from PNRR MUR project $PE0000023$-NQSTI and 
PNRR MUR project CN $00000013$ -ICSC.
T.C. acknowledges the support of PL-Grid Infrastructure for providing high-performance computing facility for a part of the numerical simulations reported here.
\end{acknowledgments}

\bibliographystyle{apsrev4-2}
\bibliography{bibliography}

\begin{thebibliography}{70}%
\makeatletter
\providecommand \@ifxundefined [1]{%
 \@ifx{#1\undefined}
}%
\providecommand \@ifnum [1]{%
 \ifnum #1\expandafter \@firstoftwo
 \else \expandafter \@secondoftwo
 \fi
}%
\providecommand \@ifx [1]{%
 \ifx #1\expandafter \@firstoftwo
 \else \expandafter \@secondoftwo
 \fi
}%
\providecommand \natexlab [1]{#1}%
\providecommand \enquote  [1]{``#1''}%
\providecommand \bibnamefont  [1]{#1}%
\providecommand \bibfnamefont [1]{#1}%
\providecommand \citenamefont [1]{#1}%
\providecommand \href@noop [0]{\@secondoftwo}%
\providecommand \href [0]{\begingroup \@sanitize@url \@href}%
\providecommand \@href[1]{\@@startlink{#1}\@@href}%
\providecommand \@@href[1]{\endgroup#1\@@endlink}%
\providecommand \@sanitize@url [0]{\catcode `\\12\catcode `\$12\catcode
  `\&12\catcode `\#12\catcode `\^12\catcode `\_12\catcode `\%12\relax}%
\providecommand \@@startlink[1]{}%
\providecommand \@@endlink[0]{}%
\providecommand \url  [0]{\begingroup\@sanitize@url \@url }%
\providecommand \@url [1]{\endgroup\@href {#1}{\urlprefix }}%
\providecommand \urlprefix  [0]{URL }%
\providecommand \Eprint [0]{\href }%
\providecommand \doibase [0]{https://doi.org/}%
\providecommand \selectlanguage [0]{\@gobble}%
\providecommand \bibinfo  [0]{\@secondoftwo}%
\providecommand \bibfield  [0]{\@secondoftwo}%
\providecommand \translation [1]{[#1]}%
\providecommand \BibitemOpen [0]{}%
\providecommand \bibitemStop [0]{}%
\providecommand \bibitemNoStop [0]{.\EOS\space}%
\providecommand \EOS [0]{\spacefactor3000\relax}%
\providecommand \BibitemShut  [1]{\csname bibitem#1\endcsname}%
\let\auto@bib@innerbib\@empty
%</preamble>
\bibitem [{\citenamefont {Shor}(1997)}]{shor1999polynomial}%
  \BibitemOpen
  \bibfield  {author} {\bibinfo {author} {\bibfnamefont {P.~W.}\ \bibnamefont
  {Shor}},\ }\href {https://doi.org/10.1137/s0097539795293172} {\bibfield
  {journal} {\bibinfo  {journal} {{SIAM} Journal on Computing}\ }\textbf
  {\bibinfo {volume} {26}},\ \bibinfo {pages} {1484} (\bibinfo {year}
  {1997})}\BibitemShut {NoStop}%
\bibitem [{\citenamefont {Feynman}(1986)}]{feynman1986quantum}%
  \BibitemOpen
  \bibfield  {author} {\bibinfo {author} {\bibfnamefont {R.~P.}\ \bibnamefont
  {Feynman}},\ }\href {https://doi.org/10.1007/bf01886518} {\bibfield
  {journal} {\bibinfo  {journal} {Foundations of Physics}\ }\textbf {\bibinfo
  {volume} {16}},\ \bibinfo {pages} {507} (\bibinfo {year} {1986})}\BibitemShut
  {NoStop}%
\bibitem [{\citenamefont {Lloyd}(1996)}]{Lloyd1996simulators}%
  \BibitemOpen
  \bibfield  {author} {\bibinfo {author} {\bibfnamefont {S.}~\bibnamefont
  {Lloyd}},\ }\href {https://doi.org/10.1126/science.273.5278.1073} {\bibfield
  {journal} {\bibinfo  {journal} {Science}\ }\textbf {\bibinfo {volume}
  {273}},\ \bibinfo {pages} {1073} (\bibinfo {year} {1996})}\BibitemShut
  {NoStop}%
\bibitem [{\citenamefont {Amico}\ \emph {et~al.}(2008)\citenamefont {Amico},
  \citenamefont {Fazio}, \citenamefont {Osterloh},\ and\ \citenamefont
  {Vedral}}]{amico2008entanglement}%
  \BibitemOpen
  \bibfield  {author} {\bibinfo {author} {\bibfnamefont {L.}~\bibnamefont
  {Amico}}, \bibinfo {author} {\bibfnamefont {R.}~\bibnamefont {Fazio}},
  \bibinfo {author} {\bibfnamefont {A.}~\bibnamefont {Osterloh}},\ and\
  \bibinfo {author} {\bibfnamefont {V.}~\bibnamefont {Vedral}},\ }\href
  {https://doi.org/10.1103/RevModPhys.80.517} {\bibfield  {journal} {\bibinfo
  {journal} {Rev. Mod. Phys.}\ }\textbf {\bibinfo {volume} {80}},\ \bibinfo
  {pages} {517} (\bibinfo {year} {2008})}\BibitemShut {NoStop}%
\bibitem [{\citenamefont {Eisert}\ \emph {et~al.}(2010)\citenamefont {Eisert},
  \citenamefont {Cramer},\ and\ \citenamefont {Plenio}}]{eisert2010colloquium}%
  \BibitemOpen
  \bibfield  {author} {\bibinfo {author} {\bibfnamefont {J.}~\bibnamefont
  {Eisert}}, \bibinfo {author} {\bibfnamefont {M.}~\bibnamefont {Cramer}},\
  and\ \bibinfo {author} {\bibfnamefont {M.~B.}\ \bibnamefont {Plenio}},\
  }\href {https://doi.org/10.1103/RevModPhys.82.277} {\bibfield  {journal}
  {\bibinfo  {journal} {Rev. Mod. Phys.}\ }\textbf {\bibinfo {volume} {82}},\
  \bibinfo {pages} {277} (\bibinfo {year} {2010})}\BibitemShut {NoStop}%
\bibitem [{\citenamefont {Cirac}\ and\ \citenamefont
  {Zoller}(2012)}]{cirac2012goals}%
  \BibitemOpen
  \bibfield  {author} {\bibinfo {author} {\bibfnamefont {J.~I.}\ \bibnamefont
  {Cirac}}\ and\ \bibinfo {author} {\bibfnamefont {P.}~\bibnamefont {Zoller}},\
  }\href {https://doi.org/10.1038/nphys2275} {\bibfield  {journal} {\bibinfo
  {journal} {Nature Physics}\ }\textbf {\bibinfo {volume} {8}},\ \bibinfo
  {pages} {264} (\bibinfo {year} {2012})}\BibitemShut {NoStop}%
\bibitem [{\citenamefont {Bloch}\ \emph {et~al.}(2012)\citenamefont {Bloch},
  \citenamefont {Dalibard},\ and\ \citenamefont
  {Nascimb{\`{e}}ne}}]{bloch2012quantum}%
  \BibitemOpen
  \bibfield  {author} {\bibinfo {author} {\bibfnamefont {I.}~\bibnamefont
  {Bloch}}, \bibinfo {author} {\bibfnamefont {J.}~\bibnamefont {Dalibard}},\
  and\ \bibinfo {author} {\bibfnamefont {S.}~\bibnamefont {Nascimb{\`{e}}ne}},\
  }\href {https://doi.org/10.1038/nphys2259} {\bibfield  {journal} {\bibinfo
  {journal} {Nature Physics}\ }\textbf {\bibinfo {volume} {8}},\ \bibinfo
  {pages} {267} (\bibinfo {year} {2012})}\BibitemShut {NoStop}%
\bibitem [{\citenamefont {Aspuru-Guzik}\ and\ \citenamefont
  {Walther}(2012)}]{aspuru2012photonic}%
  \BibitemOpen
  \bibfield  {author} {\bibinfo {author} {\bibfnamefont {A.}~\bibnamefont
  {Aspuru-Guzik}}\ and\ \bibinfo {author} {\bibfnamefont {P.}~\bibnamefont
  {Walther}},\ }\href {https://doi.org/10.1038/nphys2253} {\bibfield  {journal}
  {\bibinfo  {journal} {Nature Physics}\ }\textbf {\bibinfo {volume} {8}},\
  \bibinfo {pages} {285} (\bibinfo {year} {2012})}\BibitemShut {NoStop}%
\bibitem [{\citenamefont {Houck}\ \emph {et~al.}(2012)\citenamefont {Houck},
  \citenamefont {T\"{u}reci},\ and\ \citenamefont {Koch}}]{houck2012chip}%
  \BibitemOpen
  \bibfield  {author} {\bibinfo {author} {\bibfnamefont {A.~A.}\ \bibnamefont
  {Houck}}, \bibinfo {author} {\bibfnamefont {H.~E.}\ \bibnamefont
  {T\"{u}reci}},\ and\ \bibinfo {author} {\bibfnamefont {J.}~\bibnamefont
  {Koch}},\ }\href {https://doi.org/10.1038/nphys2251} {\bibfield  {journal}
  {\bibinfo  {journal} {Nature Physics}\ }\textbf {\bibinfo {volume} {8}},\
  \bibinfo {pages} {292} (\bibinfo {year} {2012})}\BibitemShut {NoStop}%
\bibitem [{\citenamefont {Vandersypen}\ and\ \citenamefont
  {Chuang}(2005)}]{vandersypen2005nmr}%
  \BibitemOpen
  \bibfield  {author} {\bibinfo {author} {\bibfnamefont {L.~M.~K.}\
  \bibnamefont {Vandersypen}}\ and\ \bibinfo {author} {\bibfnamefont {I.~L.}\
  \bibnamefont {Chuang}},\ }\href {https://doi.org/10.1103/RevModPhys.76.1037}
  {\bibfield  {journal} {\bibinfo  {journal} {Rev. Mod. Phys.}\ }\textbf
  {\bibinfo {volume} {76}},\ \bibinfo {pages} {1037} (\bibinfo {year}
  {2005})}\BibitemShut {NoStop}%
\bibitem [{\citenamefont {Plenio}\ and\ \citenamefont
  {Virmani}(2014)}]{plenio2014introduction}%
  \BibitemOpen
  \bibfield  {author} {\bibinfo {author} {\bibfnamefont {M.~B.}\ \bibnamefont
  {Plenio}}\ and\ \bibinfo {author} {\bibfnamefont {S.~S.}\ \bibnamefont
  {Virmani}},\ }in\ \href {https://doi.org/10.1007/978-3-319-04063-9_8} {\emph
  {\bibinfo {booktitle} {Quantum Information and Coherence}}}\ (\bibinfo
  {publisher} {Springer International Publishing},\ \bibinfo {year} {2014})\
  pp.\ \bibinfo {pages} {173--209}\BibitemShut {NoStop}%
\bibitem [{\citenamefont {Chamon}\ \emph {et~al.}(2022)\citenamefont {Chamon},
  \citenamefont {Mucciolo},\ and\ \citenamefont
  {Ruckenstein}}]{chamon_quantum_2022}%
  \BibitemOpen
  \bibfield  {author} {\bibinfo {author} {\bibfnamefont {C.}~\bibnamefont
  {Chamon}}, \bibinfo {author} {\bibfnamefont {E.~R.}\ \bibnamefont
  {Mucciolo}},\ and\ \bibinfo {author} {\bibfnamefont {A.~E.}\ \bibnamefont
  {Ruckenstein}},\ }\href {https://doi.org/10.1016/j.aop.2022.169086}
  {\bibfield  {journal} {\bibinfo  {journal} {Annals of Physics}\ }\textbf
  {\bibinfo {volume} {446}},\ \bibinfo {pages} {169086} (\bibinfo {year}
  {2022})}\BibitemShut {NoStop}%
\bibitem [{\citenamefont {Vidal}(2003)}]{vidal2003efficient}%
  \BibitemOpen
  \bibfield  {author} {\bibinfo {author} {\bibfnamefont {G.}~\bibnamefont
  {Vidal}},\ }\href {https://doi.org/10.1103/PhysRevLett.91.147902} {\bibfield
  {journal} {\bibinfo  {journal} {Phys. Rev. Lett.}\ }\textbf {\bibinfo
  {volume} {91}},\ \bibinfo {pages} {147902} (\bibinfo {year}
  {2003})}\BibitemShut {NoStop}%
\bibitem [{\citenamefont {Van~den Nest}\ \emph {et~al.}(2007)\citenamefont
  {Van~den Nest}, \citenamefont {D\"ur}, \citenamefont {Vidal},\ and\
  \citenamefont {Briegel}}]{van2007classical}%
  \BibitemOpen
  \bibfield  {author} {\bibinfo {author} {\bibfnamefont {M.}~\bibnamefont
  {Van~den Nest}}, \bibinfo {author} {\bibfnamefont {W.}~\bibnamefont {D\"ur}},
  \bibinfo {author} {\bibfnamefont {G.}~\bibnamefont {Vidal}},\ and\ \bibinfo
  {author} {\bibfnamefont {H.~J.}\ \bibnamefont {Briegel}},\ }\href
  {https://doi.org/10.1103/PhysRevA.75.012337} {\bibfield  {journal} {\bibinfo
  {journal} {Phys. Rev. A}\ }\textbf {\bibinfo {volume} {75}},\ \bibinfo
  {pages} {012337} (\bibinfo {year} {2007})}\BibitemShut {NoStop}%
\bibitem [{\citenamefont {Van~den Nest}\ \emph {et~al.}(2008)\citenamefont
  {Van~den Nest}, \citenamefont {D\"ur},\ and\ \citenamefont
  {Briegel}}]{van2008completeness}%
  \BibitemOpen
  \bibfield  {author} {\bibinfo {author} {\bibfnamefont {M.}~\bibnamefont
  {Van~den Nest}}, \bibinfo {author} {\bibfnamefont {W.}~\bibnamefont
  {D\"ur}},\ and\ \bibinfo {author} {\bibfnamefont {H.~J.}\ \bibnamefont
  {Briegel}},\ }\href {https://doi.org/10.1103/PhysRevLett.100.110501}
  {\bibfield  {journal} {\bibinfo  {journal} {Phys. Rev. Lett.}\ }\textbf
  {\bibinfo {volume} {100}},\ \bibinfo {pages} {110501} (\bibinfo {year}
  {2008})}\BibitemShut {NoStop}%
\bibitem [{\citenamefont {Van~den Nest}(2013)}]{van2013universal}%
  \BibitemOpen
  \bibfield  {author} {\bibinfo {author} {\bibfnamefont {M.}~\bibnamefont
  {Van~den Nest}},\ }\href {https://doi.org/10.1103/PhysRevLett.110.060504}
  {\bibfield  {journal} {\bibinfo  {journal} {Phys. Rev. Lett.}\ }\textbf
  {\bibinfo {volume} {110}},\ \bibinfo {pages} {060504} (\bibinfo {year}
  {2013})}\BibitemShut {NoStop}%
\bibitem [{\citenamefont {las Cuevas}\ \emph {et~al.}(2009)\citenamefont {las
  Cuevas}, \citenamefont {D\"{u}r}, \citenamefont {den Nest},\ and\
  \citenamefont {Briegel}}]{de2009completeness}%
  \BibitemOpen
  \bibfield  {author} {\bibinfo {author} {\bibfnamefont {G.~D.}\ \bibnamefont
  {las Cuevas}}, \bibinfo {author} {\bibfnamefont {W.}~\bibnamefont {D\"{u}r}},
  \bibinfo {author} {\bibfnamefont {M.~V.}\ \bibnamefont {den Nest}},\ and\
  \bibinfo {author} {\bibfnamefont {H.~J.}\ \bibnamefont {Briegel}},\ }\href
  {https://doi.org/10.1088/1742-5468/2009/07/p07001} {\bibfield  {journal}
  {\bibinfo  {journal} {Journal of Statistical Mechanics: Theory and
  Experiment}\ }\textbf {\bibinfo {volume} {2009}},\ \bibinfo {pages} {P07001}
  (\bibinfo {year} {2009})}\BibitemShut {NoStop}%
\bibitem [{\citenamefont {Nielsen}\ and\ \citenamefont
  {Chuang}(2012)}]{nielsen2002quantum}%
  \BibitemOpen
  \bibfield  {author} {\bibinfo {author} {\bibfnamefont {M.~A.}\ \bibnamefont
  {Nielsen}}\ and\ \bibinfo {author} {\bibfnamefont {I.~L.}\ \bibnamefont
  {Chuang}},\ }\href {https://doi.org/10.1017/cbo9780511976667} {\emph
  {\bibinfo {title} {Quantum Computation and Quantum Information}}}\ (\bibinfo
  {publisher} {Cambridge University Press},\ \bibinfo {year}
  {2012})\BibitemShut {NoStop}%
\bibitem [{\citenamefont {Gottesman}(1997)}]{gottesman1997stabilizer}%
  \BibitemOpen
  \bibfield  {author} {\bibinfo {author} {\bibfnamefont {D.}~\bibnamefont
  {Gottesman}},\ }\href@noop {} {\bibinfo {title} {{Stabilizer Codes and
  Quantum Error Correction}}} (\bibinfo {year} {1997}),\ \Eprint
  {https://arxiv.org/abs/arXiv:quant-ph/9705052} {arXiv:quant-ph/9705052}
  \BibitemShut {NoStop}%
\bibitem [{\citenamefont
  {Gottesman}(1998{\natexlab{a}})}]{gottesman1998theory}%
  \BibitemOpen
  \bibfield  {author} {\bibinfo {author} {\bibfnamefont {D.}~\bibnamefont
  {Gottesman}},\ }\href {https://doi.org/10.1103/PhysRevA.57.127} {\bibfield
  {journal} {\bibinfo  {journal} {Phys. Rev. A}\ }\textbf {\bibinfo {volume}
  {57}},\ \bibinfo {pages} {127} (\bibinfo {year}
  {1998}{\natexlab{a}})}\BibitemShut {NoStop}%
\bibitem [{\citenamefont {Aaronson}\ and\ \citenamefont
  {Gottesman}(2004)}]{aaronson2004improved}%
  \BibitemOpen
  \bibfield  {author} {\bibinfo {author} {\bibfnamefont {S.}~\bibnamefont
  {Aaronson}}\ and\ \bibinfo {author} {\bibfnamefont {D.}~\bibnamefont
  {Gottesman}},\ }\href {https://doi.org/10.1103/PhysRevA.70.052328} {\bibfield
   {journal} {\bibinfo  {journal} {Phys. Rev. A}\ }\textbf {\bibinfo {volume}
  {70}},\ \bibinfo {pages} {052328} (\bibinfo {year} {2004})}\BibitemShut
  {NoStop}%
\bibitem [{\citenamefont {Veitch}\ \emph {et~al.}(2014)\citenamefont {Veitch},
  \citenamefont {Mousavian}, \citenamefont {Gottesman},\ and\ \citenamefont
  {Emerson}}]{veitch2014resource}%
  \BibitemOpen
  \bibfield  {author} {\bibinfo {author} {\bibfnamefont {V.}~\bibnamefont
  {Veitch}}, \bibinfo {author} {\bibfnamefont {S.~A.~H.}\ \bibnamefont
  {Mousavian}}, \bibinfo {author} {\bibfnamefont {D.}~\bibnamefont
  {Gottesman}},\ and\ \bibinfo {author} {\bibfnamefont {J.}~\bibnamefont
  {Emerson}},\ }\href {https://doi.org/10.1088/1367-2630/16/1/013009}
  {\bibfield  {journal} {\bibinfo  {journal} {New Journal of Physics}\ }\textbf
  {\bibinfo {volume} {16}},\ \bibinfo {pages} {013009} (\bibinfo {year}
  {2014})}\BibitemShut {NoStop}%
\bibitem [{\citenamefont {Chamon}\ \emph {et~al.}(2014)\citenamefont {Chamon},
  \citenamefont {Hamma},\ and\ \citenamefont
  {Mucciolo}}]{chamon2014EmergentIrreversibilityEntanglement}%
  \BibitemOpen
  \bibfield  {author} {\bibinfo {author} {\bibfnamefont {C.}~\bibnamefont
  {Chamon}}, \bibinfo {author} {\bibfnamefont {A.}~\bibnamefont {Hamma}},\ and\
  \bibinfo {author} {\bibfnamefont {E.~R.}\ \bibnamefont {Mucciolo}},\ }\href
  {https://doi.org/10.1103/PhysRevLett.112.240501} {\bibfield  {journal}
  {\bibinfo  {journal} {Phys. Rev. Lett.}\ }\textbf {\bibinfo {volume} {112}},\
  \bibinfo {pages} {240501} (\bibinfo {year} {2014})}\BibitemShut {NoStop}%
\bibitem [{\citenamefont {Shaffer}\ \emph {et~al.}(2014)\citenamefont
  {Shaffer}, \citenamefont {Chamon}, \citenamefont {Hamma},\ and\ \citenamefont
  {Mucciolo}}]{shaffer2014IrreversibilityEntanglementSpectrum}%
  \BibitemOpen
  \bibfield  {author} {\bibinfo {author} {\bibfnamefont {D.}~\bibnamefont
  {Shaffer}}, \bibinfo {author} {\bibfnamefont {C.}~\bibnamefont {Chamon}},
  \bibinfo {author} {\bibfnamefont {A.}~\bibnamefont {Hamma}},\ and\ \bibinfo
  {author} {\bibfnamefont {E.~R.}\ \bibnamefont {Mucciolo}},\ }\href
  {https://doi.org/10.1088/1742-5468/2014/12/p12007} {\bibfield  {journal}
  {\bibinfo  {journal} {Journal of Statistical Mechanics: Theory and
  Experiment}\ }\textbf {\bibinfo {volume} {2014}},\ \bibinfo {pages} {P12007}
  (\bibinfo {year} {2014})}\BibitemShut {NoStop}%
\bibitem [{\citenamefont {Yang}\ \emph {et~al.}(2017)\citenamefont {Yang},
  \citenamefont {Hamma}, \citenamefont {Giampaolo}, \citenamefont {Mucciolo},\
  and\ \citenamefont {Chamon}}]{yang2017EntanglementComplexityQuantum}%
  \BibitemOpen
  \bibfield  {author} {\bibinfo {author} {\bibfnamefont {Z.-C.}\ \bibnamefont
  {Yang}}, \bibinfo {author} {\bibfnamefont {A.}~\bibnamefont {Hamma}},
  \bibinfo {author} {\bibfnamefont {S.~M.}\ \bibnamefont {Giampaolo}}, \bibinfo
  {author} {\bibfnamefont {E.~R.}\ \bibnamefont {Mucciolo}},\ and\ \bibinfo
  {author} {\bibfnamefont {C.}~\bibnamefont {Chamon}},\ }\href
  {https://doi.org/10.1103/PhysRevB.96.020408} {\bibfield  {journal} {\bibinfo
  {journal} {Phys. Rev. B}\ }\textbf {\bibinfo {volume} {96}},\ \bibinfo
  {pages} {020408} (\bibinfo {year} {2017})}\BibitemShut {NoStop}%
\bibitem [{\citenamefont {Zhou}\ \emph {et~al.}(2020)\citenamefont {Zhou},
  \citenamefont {Yang}, \citenamefont {Hamma},\ and\ \citenamefont
  {Chamon}}]{zhou2020SingleGateClifford}%
  \BibitemOpen
  \bibfield  {author} {\bibinfo {author} {\bibfnamefont {S.}~\bibnamefont
  {Zhou}}, \bibinfo {author} {\bibfnamefont {Z.}~\bibnamefont {Yang}}, \bibinfo
  {author} {\bibfnamefont {A.}~\bibnamefont {Hamma}},\ and\ \bibinfo {author}
  {\bibfnamefont {C.}~\bibnamefont {Chamon}},\ }\href
  {https://doi.org/10.21468/scipostphys.9.6.087} {\bibfield  {journal}
  {\bibinfo  {journal} {{SciPost} Physics}\ }\textbf {\bibinfo {volume} {9}},\
  \bibinfo {pages} {087} (\bibinfo {year} {2020})}\BibitemShut {NoStop}%
\bibitem [{\citenamefont {Leone}\ \emph
  {et~al.}(2021{\natexlab{a}})\citenamefont {Leone}, \citenamefont {Oliviero},
  \citenamefont {Zhou},\ and\ \citenamefont
  {Hamma}}]{leone2021QuantumChaosQuantum}%
  \BibitemOpen
  \bibfield  {author} {\bibinfo {author} {\bibfnamefont {L.}~\bibnamefont
  {Leone}}, \bibinfo {author} {\bibfnamefont {S.~F.~E.}\ \bibnamefont
  {Oliviero}}, \bibinfo {author} {\bibfnamefont {Y.}~\bibnamefont {Zhou}},\
  and\ \bibinfo {author} {\bibfnamefont {A.}~\bibnamefont {Hamma}},\ }\href
  {https://doi.org/10.22331/q-2021-05-04-453} {\bibfield  {journal} {\bibinfo
  {journal} {Quantum}\ }\textbf {\bibinfo {volume} {5}},\ \bibinfo {pages}
  {453} (\bibinfo {year} {2021}{\natexlab{a}})}\BibitemShut {NoStop}%
\bibitem [{\citenamefont {Oliviero}\ \emph {et~al.}(2021)\citenamefont
  {Oliviero}, \citenamefont {Leone},\ and\ \citenamefont
  {Hamma}}]{oliviero2021TransitionsEntanglementComplexity}%
  \BibitemOpen
  \bibfield  {author} {\bibinfo {author} {\bibfnamefont {S.~F.}\ \bibnamefont
  {Oliviero}}, \bibinfo {author} {\bibfnamefont {L.}~\bibnamefont {Leone}},\
  and\ \bibinfo {author} {\bibfnamefont {A.}~\bibnamefont {Hamma}},\ }\href
  {https://doi.org/10.1016/j.physleta.2021.127721} {\bibfield  {journal}
  {\bibinfo  {journal} {Physics Letters A}\ }\textbf {\bibinfo {volume}
  {418}},\ \bibinfo {pages} {127721} (\bibinfo {year} {2021})}\BibitemShut
  {NoStop}%
\bibitem [{\citenamefont {True}\ and\ \citenamefont
  {Hamma}(2022)}]{true2022TransitionsEntanglementComplexity}%
  \BibitemOpen
  \bibfield  {author} {\bibinfo {author} {\bibfnamefont {S.}~\bibnamefont
  {True}}\ and\ \bibinfo {author} {\bibfnamefont {A.}~\bibnamefont {Hamma}},\
  }\href {https://doi.org/10.22331/q-2022-09-22-818} {\bibfield  {journal}
  {\bibinfo  {journal} {Quantum}\ }\textbf {\bibinfo {volume} {6}},\ \bibinfo
  {pages} {818} (\bibinfo {year} {2022})}\BibitemShut {NoStop}%
\bibitem [{\citenamefont {Piemontese}\ \emph {et~al.}(2022)\citenamefont
  {Piemontese}, \citenamefont {Roscilde},\ and\ \citenamefont
  {Hamma}}]{piemontese2022EntanglementComplexityRokhsarKivelsonsign}%
  \BibitemOpen
  \bibfield  {author} {\bibinfo {author} {\bibfnamefont {S.}~\bibnamefont
  {Piemontese}}, \bibinfo {author} {\bibfnamefont {T.}~\bibnamefont
  {Roscilde}},\ and\ \bibinfo {author} {\bibfnamefont {A.}~\bibnamefont
  {Hamma}},\ }\href@noop {} {\bibinfo {title} {{Entanglement complexity of the
  Rokhsar-Kivelson-sign wavefunctions}}} (\bibinfo {year} {2022}),\ \Eprint
  {https://arxiv.org/abs/arXiv:2211.01428} {arXiv:2211.01428} \BibitemShut
  {NoStop}%
\bibitem [{\citenamefont {Leone}\ \emph
  {et~al.}(2021{\natexlab{b}})\citenamefont {Leone}, \citenamefont {Oliviero},\
  and\ \citenamefont {Hamma}}]{leone2020isospectral}%
  \BibitemOpen
  \bibfield  {author} {\bibinfo {author} {\bibfnamefont {L.}~\bibnamefont
  {Leone}}, \bibinfo {author} {\bibfnamefont {S.~F.~E.}\ \bibnamefont
  {Oliviero}},\ and\ \bibinfo {author} {\bibfnamefont {A.}~\bibnamefont
  {Hamma}},\ }\href {https://doi.org/10.3390/e23081073} {\bibfield  {journal}
  {\bibinfo  {journal} {Entropy}\ }\textbf {\bibinfo {volume} {23}},\ \bibinfo
  {pages} {1073} (\bibinfo {year} {2021}{\natexlab{b}})}\BibitemShut {NoStop}%
\bibitem [{\citenamefont
  {Gottesman}(1998{\natexlab{b}})}]{gottesman1998heisenberg}%
  \BibitemOpen
  \bibfield  {author} {\bibinfo {author} {\bibfnamefont {D.}~\bibnamefont
  {Gottesman}},\ }\href@noop {} {\bibinfo {title} {{The Heisenberg
  representation of quantum computers}}} (\bibinfo {year}
  {1998}{\natexlab{b}}),\ \Eprint
  {https://arxiv.org/abs/arXiv:quant-ph/9807006} {arXiv:quant-ph/9807006}
  \BibitemShut {NoStop}%
\bibitem [{\citenamefont {Bravyi}\ and\ \citenamefont
  {Kitaev}(2005)}]{bravyi2005UniversalQuantumComputation}%
  \BibitemOpen
  \bibfield  {author} {\bibinfo {author} {\bibfnamefont {S.}~\bibnamefont
  {Bravyi}}\ and\ \bibinfo {author} {\bibfnamefont {A.}~\bibnamefont
  {Kitaev}},\ }\href {https://doi.org/10.1103/PhysRevA.71.022316} {\bibfield
  {journal} {\bibinfo  {journal} {Phys. Rev. A}\ }\textbf {\bibinfo {volume}
  {71}},\ \bibinfo {pages} {022316} (\bibinfo {year} {2005})}\BibitemShut
  {NoStop}%
\bibitem [{\citenamefont {Campbell}\ \emph {et~al.}(2017)\citenamefont
  {Campbell}, \citenamefont {Terhal},\ and\ \citenamefont
  {Vuillot}}]{campbell2017roads}%
  \BibitemOpen
  \bibfield  {author} {\bibinfo {author} {\bibfnamefont {E.~T.}\ \bibnamefont
  {Campbell}}, \bibinfo {author} {\bibfnamefont {B.~M.}\ \bibnamefont
  {Terhal}},\ and\ \bibinfo {author} {\bibfnamefont {C.}~\bibnamefont
  {Vuillot}},\ }\href {https://doi.org/10.1038/nature23460} {\bibfield
  {journal} {\bibinfo  {journal} {Nature}\ }\textbf {\bibinfo {volume} {549}},\
  \bibinfo {pages} {172} (\bibinfo {year} {2017})}\BibitemShut {NoStop}%
\bibitem [{\citenamefont {Bravyi}\ and\ \citenamefont
  {Haah}(2012)}]{bravyi2012magic}%
  \BibitemOpen
  \bibfield  {author} {\bibinfo {author} {\bibfnamefont {S.}~\bibnamefont
  {Bravyi}}\ and\ \bibinfo {author} {\bibfnamefont {J.}~\bibnamefont {Haah}},\
  }\href {https://doi.org/10.1103/PhysRevA.86.052329} {\bibfield  {journal}
  {\bibinfo  {journal} {Phys. Rev. A}\ }\textbf {\bibinfo {volume} {86}},\
  \bibinfo {pages} {052329} (\bibinfo {year} {2012})}\BibitemShut {NoStop}%
\bibitem [{\citenamefont {Bravyi}\ and\ \citenamefont
  {Gosset}(2016)}]{bravyi2016improved}%
  \BibitemOpen
  \bibfield  {author} {\bibinfo {author} {\bibfnamefont {S.}~\bibnamefont
  {Bravyi}}\ and\ \bibinfo {author} {\bibfnamefont {D.}~\bibnamefont
  {Gosset}},\ }\href {https://doi.org/10.1103/PhysRevLett.116.250501}
  {\bibfield  {journal} {\bibinfo  {journal} {Phys. Rev. Lett.}\ }\textbf
  {\bibinfo {volume} {116}},\ \bibinfo {pages} {250501} (\bibinfo {year}
  {2016})}\BibitemShut {NoStop}%
\bibitem [{\citenamefont {Bravyi}\ \emph {et~al.}(2016)\citenamefont {Bravyi},
  \citenamefont {Smith},\ and\ \citenamefont {Smolin}}]{bravyi2016trading}%
  \BibitemOpen
  \bibfield  {author} {\bibinfo {author} {\bibfnamefont {S.}~\bibnamefont
  {Bravyi}}, \bibinfo {author} {\bibfnamefont {G.}~\bibnamefont {Smith}},\ and\
  \bibinfo {author} {\bibfnamefont {J.~A.}\ \bibnamefont {Smolin}},\ }\href
  {https://doi.org/10.1103/PhysRevX.6.021043} {\bibfield  {journal} {\bibinfo
  {journal} {Phys. Rev. X}\ }\textbf {\bibinfo {volume} {6}},\ \bibinfo {pages}
  {021043} (\bibinfo {year} {2016})}\BibitemShut {NoStop}%
\bibitem [{\citenamefont {Garcia}\ \emph {et~al.}(2014)\citenamefont {Garcia},
  \citenamefont {Markov},\ and\ \citenamefont {Cross}}]{garcia2014geometry}%
  \BibitemOpen
  \bibfield  {author} {\bibinfo {author} {\bibfnamefont {H.~J.}\ \bibnamefont
  {Garcia}}, \bibinfo {author} {\bibfnamefont {I.~L.}\ \bibnamefont {Markov}},\
  and\ \bibinfo {author} {\bibfnamefont {A.~W.}\ \bibnamefont {Cross}},\ }\href
  {https://doi.org/10.26421/qic14.7-8-9} {\bibfield  {journal} {\bibinfo
  {journal} {Quantum Information and Computation}\ }\textbf {\bibinfo {volume}
  {14}},\ \bibinfo {pages} {683} (\bibinfo {year} {2014})}\BibitemShut
  {NoStop}%
\bibitem [{\citenamefont {Wigner}(1932)}]{wigner1997quantum}%
  \BibitemOpen
  \bibfield  {author} {\bibinfo {author} {\bibfnamefont {E.}~\bibnamefont
  {Wigner}},\ }\href {https://doi.org/10.1103/PhysRev.40.749} {\bibfield
  {journal} {\bibinfo  {journal} {Phys. Rev.}\ }\textbf {\bibinfo {volume}
  {40}},\ \bibinfo {pages} {749} (\bibinfo {year} {1932})}\BibitemShut
  {NoStop}%
\bibitem [{\citenamefont {Gross}(2006)}]{gross2007non}%
  \BibitemOpen
  \bibfield  {author} {\bibinfo {author} {\bibfnamefont {D.}~\bibnamefont
  {Gross}},\ }\href {https://doi.org/10.1007/s00340-006-2510-9} {\bibfield
  {journal} {\bibinfo  {journal} {Applied Physics B}\ }\textbf {\bibinfo
  {volume} {86}},\ \bibinfo {pages} {367} (\bibinfo {year} {2006})}\BibitemShut
  {NoStop}%
\bibitem [{\citenamefont {Veitch}\ \emph {et~al.}(2012)\citenamefont {Veitch},
  \citenamefont {Ferrie}, \citenamefont {Gross},\ and\ \citenamefont
  {Emerson}}]{veitch2012negative}%
  \BibitemOpen
  \bibfield  {author} {\bibinfo {author} {\bibfnamefont {V.}~\bibnamefont
  {Veitch}}, \bibinfo {author} {\bibfnamefont {C.}~\bibnamefont {Ferrie}},
  \bibinfo {author} {\bibfnamefont {D.}~\bibnamefont {Gross}},\ and\ \bibinfo
  {author} {\bibfnamefont {J.}~\bibnamefont {Emerson}},\ }\href
  {https://doi.org/10.1088/1367-2630/14/11/113011} {\bibfield  {journal}
  {\bibinfo  {journal} {New Journal of Physics}\ }\textbf {\bibinfo {volume}
  {14}},\ \bibinfo {pages} {113011} (\bibinfo {year} {2012})}\BibitemShut
  {NoStop}%
\bibitem [{\citenamefont {Wootters}(1987)}]{wootters1987wigner}%
  \BibitemOpen
  \bibfield  {author} {\bibinfo {author} {\bibfnamefont {W.~K.}\ \bibnamefont
  {Wootters}},\ }\href {https://doi.org/10.1016/0003-4916(87)90176-x}
  {\bibfield  {journal} {\bibinfo  {journal} {Annals of Physics}\ }\textbf
  {\bibinfo {volume} {176}},\ \bibinfo {pages} {1} (\bibinfo {year}
  {1987})}\BibitemShut {NoStop}%
\bibitem [{\citenamefont {Hudson}(1974)}]{hudson1974wigner}%
  \BibitemOpen
  \bibfield  {author} {\bibinfo {author} {\bibfnamefont {R.}~\bibnamefont
  {Hudson}},\ }\href {https://doi.org/10.1016/0034-4877(74)90007-x} {\bibfield
  {journal} {\bibinfo  {journal} {Reports on Mathematical Physics}\ }\textbf
  {\bibinfo {volume} {6}},\ \bibinfo {pages} {249} (\bibinfo {year}
  {1974})}\BibitemShut {NoStop}%
\bibitem [{\citenamefont {Bu\ifmmode~\check{z}\else \v{z}\fi{}ek}\ \emph
  {et~al.}(1992)\citenamefont {Bu\ifmmode~\check{z}\else \v{z}\fi{}ek},
  \citenamefont {Vidiella-Barranco},\ and\ \citenamefont
  {Knight}}]{buvzek1992superpositions}%
  \BibitemOpen
  \bibfield  {author} {\bibinfo {author} {\bibfnamefont {V.}~\bibnamefont
  {Bu\ifmmode~\check{z}\else \v{z}\fi{}ek}}, \bibinfo {author} {\bibfnamefont
  {A.}~\bibnamefont {Vidiella-Barranco}},\ and\ \bibinfo {author}
  {\bibfnamefont {P.~L.}\ \bibnamefont {Knight}},\ }\href
  {https://doi.org/10.1103/PhysRevA.45.6570} {\bibfield  {journal} {\bibinfo
  {journal} {Phys. Rev. A}\ }\textbf {\bibinfo {volume} {45}},\ \bibinfo
  {pages} {6570} (\bibinfo {year} {1992})}\BibitemShut {NoStop}%
\bibitem [{\citenamefont {Bravyi}\ \emph {et~al.}(2019)\citenamefont {Bravyi},
  \citenamefont {Browne}, \citenamefont {Calpin}, \citenamefont {Campbell},
  \citenamefont {Gosset},\ and\ \citenamefont {Howard}}]{bravyi2019simulation}%
  \BibitemOpen
  \bibfield  {author} {\bibinfo {author} {\bibfnamefont {S.}~\bibnamefont
  {Bravyi}}, \bibinfo {author} {\bibfnamefont {D.}~\bibnamefont {Browne}},
  \bibinfo {author} {\bibfnamefont {P.}~\bibnamefont {Calpin}}, \bibinfo
  {author} {\bibfnamefont {E.}~\bibnamefont {Campbell}}, \bibinfo {author}
  {\bibfnamefont {D.}~\bibnamefont {Gosset}},\ and\ \bibinfo {author}
  {\bibfnamefont {M.}~\bibnamefont {Howard}},\ }\href
  {https://doi.org/10.22331/q-2019-09-02-181} {\bibfield  {journal} {\bibinfo
  {journal} {Quantum}\ }\textbf {\bibinfo {volume} {3}},\ \bibinfo {pages}
  {181} (\bibinfo {year} {2019})}\BibitemShut {NoStop}%
\bibitem [{\citenamefont {Howard}\ and\ \citenamefont
  {Campbell}(2017)}]{howard2017robustness}%
  \BibitemOpen
  \bibfield  {author} {\bibinfo {author} {\bibfnamefont {M.}~\bibnamefont
  {Howard}}\ and\ \bibinfo {author} {\bibfnamefont {E.}~\bibnamefont
  {Campbell}},\ }\href {https://doi.org/10.1103/PhysRevLett.118.090501}
  {\bibfield  {journal} {\bibinfo  {journal} {Phys. Rev. Lett.}\ }\textbf
  {\bibinfo {volume} {118}},\ \bibinfo {pages} {090501} (\bibinfo {year}
  {2017})}\BibitemShut {NoStop}%
\bibitem [{\citenamefont {Heinrich}\ and\ \citenamefont
  {Gross}(2019)}]{heinrich2019robustness}%
  \BibitemOpen
  \bibfield  {author} {\bibinfo {author} {\bibfnamefont {M.}~\bibnamefont
  {Heinrich}}\ and\ \bibinfo {author} {\bibfnamefont {D.}~\bibnamefont
  {Gross}},\ }\href {https://doi.org/10.22331/q-2019-04-08-132} {\bibfield
  {journal} {\bibinfo  {journal} {Quantum}\ }\textbf {\bibinfo {volume} {3}},\
  \bibinfo {pages} {132} (\bibinfo {year} {2019})}\BibitemShut {NoStop}%
\bibitem [{\citenamefont {Wang}\ \emph {et~al.}(2020)\citenamefont {Wang},
  \citenamefont {Wilde},\ and\ \citenamefont {Su}}]{wang2020efficiently}%
  \BibitemOpen
  \bibfield  {author} {\bibinfo {author} {\bibfnamefont {X.}~\bibnamefont
  {Wang}}, \bibinfo {author} {\bibfnamefont {M.~M.}\ \bibnamefont {Wilde}},\
  and\ \bibinfo {author} {\bibfnamefont {Y.}~\bibnamefont {Su}},\ }\href
  {https://doi.org/10.1103/PhysRevLett.124.090505} {\bibfield  {journal}
  {\bibinfo  {journal} {Phys. Rev. Lett.}\ }\textbf {\bibinfo {volume} {124}},\
  \bibinfo {pages} {090505} (\bibinfo {year} {2020})}\BibitemShut {NoStop}%
\bibitem [{\citenamefont {Heimendahl}\ \emph {et~al.}(2021)\citenamefont
  {Heimendahl}, \citenamefont {Montealegre-Mora}, \citenamefont {Vallentin},\
  and\ \citenamefont {Gross}}]{heimendahl2021stabilizer}%
  \BibitemOpen
  \bibfield  {author} {\bibinfo {author} {\bibfnamefont {A.}~\bibnamefont
  {Heimendahl}}, \bibinfo {author} {\bibfnamefont {F.}~\bibnamefont
  {Montealegre-Mora}}, \bibinfo {author} {\bibfnamefont {F.}~\bibnamefont
  {Vallentin}},\ and\ \bibinfo {author} {\bibfnamefont {D.}~\bibnamefont
  {Gross}},\ }\href {https://doi.org/10.22331/q-2021-02-24-400} {\bibfield
  {journal} {\bibinfo  {journal} {Quantum}\ }\textbf {\bibinfo {volume} {5}},\
  \bibinfo {pages} {400} (\bibinfo {year} {2021})}\BibitemShut {NoStop}%
\bibitem [{\citenamefont {Leone}\ \emph {et~al.}(2022)\citenamefont {Leone},
  \citenamefont {Oliviero},\ and\ \citenamefont {Hamma}}]{leone2022stabilizer}%
  \BibitemOpen
  \bibfield  {author} {\bibinfo {author} {\bibfnamefont {L.}~\bibnamefont
  {Leone}}, \bibinfo {author} {\bibfnamefont {S.~F.~E.}\ \bibnamefont
  {Oliviero}},\ and\ \bibinfo {author} {\bibfnamefont {A.}~\bibnamefont
  {Hamma}},\ }\href {https://doi.org/10.1103/PhysRevLett.128.050402} {\bibfield
   {journal} {\bibinfo  {journal} {Phys. Rev. Lett.}\ }\textbf {\bibinfo
  {volume} {128}},\ \bibinfo {pages} {050402} (\bibinfo {year}
  {2022})}\BibitemShut {NoStop}%
\bibitem [{\citenamefont {Oliviero}\ \emph
  {et~al.}(2022{\natexlab{a}})\citenamefont {Oliviero}, \citenamefont {Leone},\
  and\ \citenamefont {Hamma}}]{oliviero2022MagicstateResourceTheory}%
  \BibitemOpen
  \bibfield  {author} {\bibinfo {author} {\bibfnamefont {S.~F.~E.}\
  \bibnamefont {Oliviero}}, \bibinfo {author} {\bibfnamefont {L.}~\bibnamefont
  {Leone}},\ and\ \bibinfo {author} {\bibfnamefont {A.}~\bibnamefont {Hamma}},\
  }\href {https://doi.org/10.1103/PhysRevA.106.042426} {\bibfield  {journal}
  {\bibinfo  {journal} {Phys. Rev. A}\ }\textbf {\bibinfo {volume} {106}},\
  \bibinfo {pages} {042426} (\bibinfo {year} {2022}{\natexlab{a}})}\BibitemShut
  {NoStop}%
\bibitem [{\citenamefont {Haug}\ and\ \citenamefont
  {Piroli}(2023{\natexlab{a}})}]{haug2023quantifying}%
  \BibitemOpen
  \bibfield  {author} {\bibinfo {author} {\bibfnamefont {T.}~\bibnamefont
  {Haug}}\ and\ \bibinfo {author} {\bibfnamefont {L.}~\bibnamefont {Piroli}},\
  }\href {https://doi.org/10.1103/PhysRevB.107.035148} {\bibfield  {journal}
  {\bibinfo  {journal} {Phys. Rev. B}\ }\textbf {\bibinfo {volume} {107}},\
  \bibinfo {pages} {035148} (\bibinfo {year} {2023}{\natexlab{a}})}\BibitemShut
  {NoStop}%
\bibitem [{\citenamefont {Haug}\ and\ \citenamefont
  {Piroli}(2023{\natexlab{b}})}]{haug2023stabilizer}%
  \BibitemOpen
  \bibfield  {author} {\bibinfo {author} {\bibfnamefont {T.}~\bibnamefont
  {Haug}}\ and\ \bibinfo {author} {\bibfnamefont {L.}~\bibnamefont {Piroli}},\
  }\href@noop {} {\bibinfo {title} {{Stabilizer entropies and nonstabilizerness
  monotones}}} (\bibinfo {year} {2023}{\natexlab{b}}),\ \Eprint
  {https://arxiv.org/abs/arXiv:2303.10152} {arXiv:2303.10152} \BibitemShut
  {NoStop}%
\bibitem [{\citenamefont {Tarabunga}\ \emph {et~al.}(2023)\citenamefont
  {Tarabunga}, \citenamefont {Tirrito}, \citenamefont {Chanda},\ and\
  \citenamefont {Dalmonte}}]{tarabunga2023manybody}%
  \BibitemOpen
  \bibfield  {author} {\bibinfo {author} {\bibfnamefont {P.~S.}\ \bibnamefont
  {Tarabunga}}, \bibinfo {author} {\bibfnamefont {E.}~\bibnamefont {Tirrito}},
  \bibinfo {author} {\bibfnamefont {T.}~\bibnamefont {Chanda}},\ and\ \bibinfo
  {author} {\bibfnamefont {M.}~\bibnamefont {Dalmonte}},\ }\href@noop {}
  {\bibinfo {title} {Many-body magic via pauli-markov chains -- from
  criticality to gauge theories}} (\bibinfo {year} {2023}),\ \Eprint
  {https://arxiv.org/abs/2305.18541} {arXiv:2305.18541 [quant-ph]} \BibitemShut
  {NoStop}%
\bibitem [{\citenamefont {Haug}\ and\ \citenamefont
  {Kim}(2023)}]{haug2022scalable}%
  \BibitemOpen
  \bibfield  {author} {\bibinfo {author} {\bibfnamefont {T.}~\bibnamefont
  {Haug}}\ and\ \bibinfo {author} {\bibfnamefont {M.}~\bibnamefont {Kim}},\
  }\href {https://doi.org/10.1103/PRXQuantum.4.010301} {\bibfield  {journal}
  {\bibinfo  {journal} {PRX Quantum}\ }\textbf {\bibinfo {volume} {4}},\
  \bibinfo {pages} {010301} (\bibinfo {year} {2023})}\BibitemShut {NoStop}%
\bibitem [{\citenamefont {Oliviero}\ \emph
  {et~al.}(2022{\natexlab{b}})\citenamefont {Oliviero}, \citenamefont {Leone},
  \citenamefont {Hamma},\ and\ \citenamefont {Lloyd}}]{oliviero2022measuring}%
  \BibitemOpen
  \bibfield  {author} {\bibinfo {author} {\bibfnamefont {S.~F.~E.}\
  \bibnamefont {Oliviero}}, \bibinfo {author} {\bibfnamefont {L.}~\bibnamefont
  {Leone}}, \bibinfo {author} {\bibfnamefont {A.}~\bibnamefont {Hamma}},\ and\
  \bibinfo {author} {\bibfnamefont {S.}~\bibnamefont {Lloyd}},\ }\href
  {https://doi.org/10.1038/s41534-022-00666-5} {\bibfield  {journal} {\bibinfo
  {journal} {npj Quantum Information}\ }\textbf {\bibinfo {volume} {8}},\
  \bibinfo {pages} {148} (\bibinfo {year} {2022}{\natexlab{b}})}\BibitemShut
  {NoStop}%
\bibitem [{\citenamefont {Leone}\ \emph
  {et~al.}(2023{\natexlab{a}})\citenamefont {Leone}, \citenamefont {Oliviero},\
  and\ \citenamefont {Hamma}}]{Leone2023nonstabilizernesshardness}%
  \BibitemOpen
  \bibfield  {author} {\bibinfo {author} {\bibfnamefont {L.}~\bibnamefont
  {Leone}}, \bibinfo {author} {\bibfnamefont {S.~F.~E.}\ \bibnamefont
  {Oliviero}},\ and\ \bibinfo {author} {\bibfnamefont {A.}~\bibnamefont
  {Hamma}},\ }\href {https://doi.org/10.1103/PhysRevA.107.022429} {\bibfield
  {journal} {\bibinfo  {journal} {Phys. Rev. A}\ }\textbf {\bibinfo {volume}
  {107}},\ \bibinfo {pages} {022429} (\bibinfo {year}
  {2023}{\natexlab{a}})}\BibitemShut {NoStop}%
\bibitem [{See()}]{SeeSupplementalMaterial}%
  \BibitemOpen
  \href@noop {} {\bibinfo {title} {See supplemental material for formal proofs,
  which includes
  refs.~\cite{zhu2017MultiqubitCliffordGroups,zhu2016CliffordGroupFails,oliviero2021TransitionsEntanglementComplexity}}}\BibitemShut
  {NoStop}%
\bibitem [{\citenamefont {Elben}\ \emph {et~al.}(2023)\citenamefont {Elben},
  \citenamefont {Flammia}, \citenamefont {Huang}, \citenamefont {Kueng},
  \citenamefont {Preskill}, \citenamefont {Vermersch},\ and\ \citenamefont
  {Zoller}}]{elben2023randomized}%
  \BibitemOpen
  \bibfield  {author} {\bibinfo {author} {\bibfnamefont {A.}~\bibnamefont
  {Elben}}, \bibinfo {author} {\bibfnamefont {S.~T.}\ \bibnamefont {Flammia}},
  \bibinfo {author} {\bibfnamefont {H.-Y.}\ \bibnamefont {Huang}}, \bibinfo
  {author} {\bibfnamefont {R.}~\bibnamefont {Kueng}}, \bibinfo {author}
  {\bibfnamefont {J.}~\bibnamefont {Preskill}}, \bibinfo {author}
  {\bibfnamefont {B.}~\bibnamefont {Vermersch}},\ and\ \bibinfo {author}
  {\bibfnamefont {P.}~\bibnamefont {Zoller}},\ }\href@noop {} {\bibfield
  {journal} {\bibinfo  {journal} {Nature Reviews Physics}\ }\textbf {\bibinfo
  {volume} {5}},\ \bibinfo {pages} {9} (\bibinfo {year} {2023})}\BibitemShut
  {NoStop}%
\bibitem [{inp()}]{inprep}%
  \BibitemOpen
  \href@noop {} {}\bibinfo {note} {A thorough discussion of the finite-size
  scaling will be found in E. Tirrito et. al., in preparation.}\BibitemShut
  {Stop}%
\bibitem [{\citenamefont {Pichler}\ \emph {et~al.}(2016)\citenamefont
  {Pichler}, \citenamefont {Zhu}, \citenamefont {Seif}, \citenamefont
  {Zoller},\ and\ \citenamefont {Hafezi}}]{pichler2016measurement}%
  \BibitemOpen
  \bibfield  {author} {\bibinfo {author} {\bibfnamefont {H.}~\bibnamefont
  {Pichler}}, \bibinfo {author} {\bibfnamefont {G.}~\bibnamefont {Zhu}},
  \bibinfo {author} {\bibfnamefont {A.}~\bibnamefont {Seif}}, \bibinfo {author}
  {\bibfnamefont {P.}~\bibnamefont {Zoller}},\ and\ \bibinfo {author}
  {\bibfnamefont {M.}~\bibnamefont {Hafezi}},\ }\href
  {https://doi.org/10.1103/PhysRevX.6.041033} {\bibfield  {journal} {\bibinfo
  {journal} {Phys. Rev. X}\ }\textbf {\bibinfo {volume} {6}},\ \bibinfo {pages}
  {041033} (\bibinfo {year} {2016})}\BibitemShut {NoStop}%
\bibitem [{\citenamefont {Johri}\ \emph {et~al.}(2017)\citenamefont {Johri},
  \citenamefont {Steiger},\ and\ \citenamefont
  {Troyer}}]{johri2017entanglement}%
  \BibitemOpen
  \bibfield  {author} {\bibinfo {author} {\bibfnamefont {S.}~\bibnamefont
  {Johri}}, \bibinfo {author} {\bibfnamefont {D.~S.}\ \bibnamefont {Steiger}},\
  and\ \bibinfo {author} {\bibfnamefont {M.}~\bibnamefont {Troyer}},\ }\href
  {https://doi.org/10.1103/PhysRevB.96.195136} {\bibfield  {journal} {\bibinfo
  {journal} {Phys. Rev. B}\ }\textbf {\bibinfo {volume} {96}},\ \bibinfo
  {pages} {195136} (\bibinfo {year} {2017})}\BibitemShut {NoStop}%
\bibitem [{\citenamefont {Choo}\ \emph {et~al.}(2018)\citenamefont {Choo},
  \citenamefont {von Keyserlingk}, \citenamefont {Regnault},\ and\
  \citenamefont {Neupert}}]{choo2018measurement}%
  \BibitemOpen
  \bibfield  {author} {\bibinfo {author} {\bibfnamefont {K.}~\bibnamefont
  {Choo}}, \bibinfo {author} {\bibfnamefont {C.~W.}\ \bibnamefont {von
  Keyserlingk}}, \bibinfo {author} {\bibfnamefont {N.}~\bibnamefont
  {Regnault}},\ and\ \bibinfo {author} {\bibfnamefont {T.}~\bibnamefont
  {Neupert}},\ }\href {https://doi.org/10.1103/PhysRevLett.121.086808}
  {\bibfield  {journal} {\bibinfo  {journal} {Phys. Rev. Lett.}\ }\textbf
  {\bibinfo {volume} {121}},\ \bibinfo {pages} {086808} (\bibinfo {year}
  {2018})}\BibitemShut {NoStop}%
\bibitem [{\citenamefont {Leone}\ \emph
  {et~al.}(2023{\natexlab{b}})\citenamefont {Leone}, \citenamefont {Oliviero},
  \citenamefont {Esposito},\ and\ \citenamefont {Hamma}}]{leone2023phase}%
  \BibitemOpen
  \bibfield  {author} {\bibinfo {author} {\bibfnamefont {L.}~\bibnamefont
  {Leone}}, \bibinfo {author} {\bibfnamefont {S.~F.~E.}\ \bibnamefont
  {Oliviero}}, \bibinfo {author} {\bibfnamefont {G.}~\bibnamefont {Esposito}},\
  and\ \bibinfo {author} {\bibfnamefont {A.}~\bibnamefont {Hamma}},\
  }\href@noop {} {\bibinfo {title} {Phase transition in stabilizer entropy and
  efficient purity estimation}} (\bibinfo {year} {2023}{\natexlab{b}}),\
  \Eprint {https://arxiv.org/abs/2302.07895} {arXiv:2302.07895 [quant-ph]}
  \BibitemShut {NoStop}%
\bibitem [{\citenamefont {Kitaev}(2003)}]{kitaev2003fault}%
  \BibitemOpen
  \bibfield  {author} {\bibinfo {author} {\bibfnamefont {A.}~\bibnamefont
  {Kitaev}},\ }\href {https://doi.org/10.1016/s0003-4916(02)00018-0} {\bibfield
   {journal} {\bibinfo  {journal} {Annals of Physics}\ }\textbf {\bibinfo
  {volume} {303}},\ \bibinfo {pages} {2} (\bibinfo {year} {2003})}\BibitemShut
  {NoStop}%
\bibitem [{\citenamefont {Dennis}\ \emph {et~al.}(2002)\citenamefont {Dennis},
  \citenamefont {Kitaev}, \citenamefont {Landahl},\ and\ \citenamefont
  {Preskill}}]{dennis2002topological}%
  \BibitemOpen
  \bibfield  {author} {\bibinfo {author} {\bibfnamefont {E.}~\bibnamefont
  {Dennis}}, \bibinfo {author} {\bibfnamefont {A.}~\bibnamefont {Kitaev}},
  \bibinfo {author} {\bibfnamefont {A.}~\bibnamefont {Landahl}},\ and\ \bibinfo
  {author} {\bibfnamefont {J.}~\bibnamefont {Preskill}},\ }\href
  {https://doi.org/10.1063/1.1499754} {\bibfield  {journal} {\bibinfo
  {journal} {Journal of Mathematical Physics}\ }\textbf {\bibinfo {volume}
  {43}},\ \bibinfo {pages} {4452} (\bibinfo {year} {2002})}\BibitemShut
  {NoStop}%
\bibitem [{\citenamefont {Raussendorf}\ \emph {et~al.}(2007)\citenamefont
  {Raussendorf}, \citenamefont {Harrington},\ and\ \citenamefont
  {Goyal}}]{raussendorf2007topological}%
  \BibitemOpen
  \bibfield  {author} {\bibinfo {author} {\bibfnamefont {R.}~\bibnamefont
  {Raussendorf}}, \bibinfo {author} {\bibfnamefont {J.}~\bibnamefont
  {Harrington}},\ and\ \bibinfo {author} {\bibfnamefont {K.}~\bibnamefont
  {Goyal}},\ }\href {https://doi.org/10.1088/1367-2630/9/6/199} {\bibfield
  {journal} {\bibinfo  {journal} {New Journal of Physics}\ }\textbf {\bibinfo
  {volume} {9}},\ \bibinfo {pages} {199} (\bibinfo {year} {2007})}\BibitemShut
  {NoStop}%
\bibitem [{\citenamefont {Fowler}\ \emph {et~al.}(2012)\citenamefont {Fowler},
  \citenamefont {Mariantoni}, \citenamefont {Martinis},\ and\ \citenamefont
  {Cleland}}]{fowler2012surface}%
  \BibitemOpen
  \bibfield  {author} {\bibinfo {author} {\bibfnamefont {A.~G.}\ \bibnamefont
  {Fowler}}, \bibinfo {author} {\bibfnamefont {M.}~\bibnamefont {Mariantoni}},
  \bibinfo {author} {\bibfnamefont {J.~M.}\ \bibnamefont {Martinis}},\ and\
  \bibinfo {author} {\bibfnamefont {A.~N.}\ \bibnamefont {Cleland}},\ }\href
  {https://doi.org/10.1103/PhysRevA.86.032324} {\bibfield  {journal} {\bibinfo
  {journal} {Phys. Rev. A}\ }\textbf {\bibinfo {volume} {86}},\ \bibinfo
  {pages} {032324} (\bibinfo {year} {2012})}\BibitemShut {NoStop}%
\bibitem [{\citenamefont {Zhu}(2017)}]{zhu2017MultiqubitCliffordGroups}%
  \BibitemOpen
  \bibfield  {author} {\bibinfo {author} {\bibfnamefont {H.}~\bibnamefont
  {Zhu}},\ }\href {https://doi.org/10.1103/PhysRevA.96.062336} {\bibfield
  {journal} {\bibinfo  {journal} {Phys. Rev. A}\ }\textbf {\bibinfo {volume}
  {96}},\ \bibinfo {pages} {062336} (\bibinfo {year} {2017})}\BibitemShut
  {NoStop}%
\bibitem [{\citenamefont {Zhu}\ \emph {et~al.}(2016)\citenamefont {Zhu},
  \citenamefont {Kueng}, \citenamefont {Grassl},\ and\ \citenamefont
  {Gross}}]{zhu2016CliffordGroupFails}%
  \BibitemOpen
  \bibfield  {author} {\bibinfo {author} {\bibfnamefont {H.}~\bibnamefont
  {Zhu}}, \bibinfo {author} {\bibfnamefont {R.}~\bibnamefont {Kueng}}, \bibinfo
  {author} {\bibfnamefont {M.}~\bibnamefont {Grassl}},\ and\ \bibinfo {author}
  {\bibfnamefont {D.}~\bibnamefont {Gross}},\ }\href@noop {} {\bibinfo {title}
  {{The Clifford group fails gracefully to be a unitary 4-design}}} (\bibinfo
  {year} {2016}),\ \Eprint {https://arxiv.org/abs/arXiv:1609.08172}
  {arXiv:1609.08172} \BibitemShut {NoStop}%
\end{thebibliography}%

    \foreach \x in {1,...,\numbersupplementpages}
    {
        \clearpage
        \includepdf[pages={\x,{}}]{\supplementfilename}
    }

\end{document}